\pdfoutput=1        

\documentclass[%
  aps,
  prd,
  nofootinbib,
  preprintnumbers,
  superscriptaddress,
  floatfix,
]{revtex4-2}

\usepackage{graphicx}
\usepackage{dcolumn}
\usepackage{bm}
\usepackage{amsmath}
\usepackage{amssymb}
\usepackage{amsthm}
\usepackage{xcolor}
\usepackage{physics}
\usepackage{tensor}
\usepackage{cases}
\usepackage{appendix}

\usepackage[%
  bookmarksnumbered,
  pdfpagelabels=true,
  plainpages=false,
  colorlinks=true,
  linkcolor=blue,
  citecolor=blue,
  urlcolor=blue,
]{hyperref}

\theoremstyle{plain}

\newcommand{\cA}{\mathcal A}

\newcommand{\cQ}{\mathcal Q}

\newcommand{\ep}{\varepsilon}

\newcommand{\be}{\begin{equation}}
\newcommand{\ee}{\end{equation}}
\newcommand{\bea}{\begin{eqnarray}}
\newcommand{\eea}{\end{eqnarray}}

\newcommand{\bml}{\begin{subequations}}
\newcommand{\eml}{\end{subequations}}

\newcommand{\bbm}{\begin{bmatrix}}
\newcommand{\ebm}{\end{bmatrix}}
\newcommand{\bvm}{\begin{vmatrix}}
\newcommand{\evm}{\end{vmatrix}}

\newcommand\m[1]{\begin{pmatrix}#1\end{pmatrix}}

\newcommand\eq[1]{\begin{equation} #1 \end{equation}}

\newcommand{\sgn}{\text{sgn}}
\newcommand{\minmod}{\text{minmod}}

\begin{document}


\title{Flux-Conservative BDNK Hydrodynamics and Shock Regularization}
\date{\today}

\author{Nicolas Clarisse}
\affiliation{Illinois Center for Advanced Studies of the Universe\\ 
Department of Physics, University of Illinois at Urbana-Champaign, 
Urbana, IL 61801, USA}
\email{nclari2@illinois.edu}

\author{Eduardo O. Pinho}
\affiliation{Departamento de F\'isica, CFM - Universidade Federal de
Santa Catarina; C.P. 476, CEP 88.040-900, Florian\'opolis, SC, Brazil}
\email{eduardo.op@posgrad.ufsc.br}

\author{Teerthal Patel}
\affiliation{Department of Mathematics, Vanderbilt University, 
Nashville, TN, 37235, USA}
\email{teerthal.patel@vanderbilt.edu}

\author{F\'abio S. Bemfica}
\altaffiliation{Also at Escola de Ci\^encias e Tecnologia, 
Universidade Federal do Rio Grande do Norte, RN, 59072-970, Natal, Brazil}
\affiliation{Department of Mathematics, Vanderbilt University, 
Nashville, TN, 37235, USA}
\email{fabio.bemfica@ect.ufrn.br}

\author{Maur\'icio Hippert}
\affiliation{Centro Brasileiro de Pesquisas F\'isicas, 
Rua Dr. Xavier Sigaud 150, 22290-180 Rio de Janeiro, Rio de Janeiro, Brazil}
\email{hippert@cbpf.br}

\author{Jorge Noronha}
\affiliation{Illinois Center for Advanced Studies of the Universe\\ 
Department of Physics, University of Illinois at Urbana-Champaign, 
Urbana, IL 61801, USA}
\email{jn0508@illinois.edu}

\begin{abstract}
We present a new, first-order, flux-conservative formulation of relativistic viscous hydrodynamics in the BDNK framework, applicable to conformal and nonconformal fluids at zero chemical potential. Focusing on the conformal case in 1+1 dimensions, we numerically solve the equations of motion for two classes of consistent initial data and assess the robustness of the resulting solutions with respect to changing the hydrodynamic frame. Our flux-conservative formulation does not exhibit spurious oscillatory structures within the regime of validity of BDNK theory (the hydrodynamic-frame–robust regime), corresponding to sufficiently small Knudsen numbers. Using this flux-conservative approach, we numerically investigate the potential formation of shocks and their fate in BDNK in 1+1D using smooth initial data known to produce shocks in the relativistic Euler equations. For the type of initial data we consider, we show that the sharp features formed in Euler are prevented by the hyperbolic viscous BDNK equations. The prevention of shock formation we observe occurs in the frame-robust regime of BDNK. This, however, does not preclude the formation of shocks in BDNK for different smooth initial data. The reliability of our results is supported by systematic numerical convergence testing, which is used to assess shock indicators, simulation performance, and the robustness of solutions for different hydrodynamic frames.
\end{abstract}

\maketitle

\section{Introduction}

Hydrodynamics captures the underlying physics associated with the dynamics of conserved quantities at long timescales and long wavelengths \cite{LandauLifshitzFluids}. Its power and generality stem from the fact that the equations of motion of hydrodynamics derive from conservation  laws which, in the case of a nonrelativistic system, correspond to the local conservation of mass, energy, and momentum. In ideal fluids, the system of equations is closed once an equation of state is specified, and local well-posedness holds --- that is, local (i.e., finite in time) solutions are guaranteed to exist and be unique for suitable initial data \cite{AnileBook}.

Away from equilibrium, further assumptions are needed to determine the dynamics of dissipative fluxes and solve the corresponding system of partial differential equations. Typically, hydrodynamics is expected to provide a good approximation in situations where there is clear separation of length scales, which justifies the systematic truncation of the constitutive relations defining the dissipative fluxes in terms of gradients of the original dynamical fields \cite{LandauLifshitzFluids}. In practice, such an approach is valid to describe phenomena where the typical scale $\ell$ associated with gradients of the fluid variables is much larger than $l_{\text{micro}}$, the microscopic length scale associated with interactions (in a gas, one may take this to be of the order of the mean free path). This defines the small Knudsen number regime, $\text{Kn} = l_{\text{micro}}/\ell \ll 1$, where the hydrodynamic description is expected to be valid \cite{LandauLifshitzFluids,Rocha:2023ilf}. This reasoning, when truncated at first order in gradients, leads to the famous Navier-Stokes equations \cite{LandauLifshitzFluids}, which describe a wide set of phenomena ranging from simple, smooth laminar flows to the complex behavior found in turbulence \cite{Frisch_1995}.

Similar arguments may be applied to investigate how relativity modifies the equations of motion of fluid dynamics  \cite{Rezzolla_Zanotti_book}. At the core of relativistic hydrodynamics is the covariant conservation of the energy-momentum tensor:
\eq{
\nabla_\mu T^{\mu\nu} = 0. 
}
In ideal relativistic hydrodynamics, the energy-momentum tensor takes the form
\be
T^{\mu\nu}_0 = \varepsilon u^\mu u^\nu + P\Delta^{\mu\nu}
\label{eq:ideal_hydro_Tmunu}
\ee
where $u^\mu$ is the fluid velocity (a future-directed, timelike 4-vector normalized as $u_\mu u^\mu = -1$), $\Delta_{\mu\nu} = g_{\mu\nu} + u_\mu u_\nu$ is the projector orthogonal to the flow (with $g_{\mu\nu}$ being the  spacetime metric), $\varepsilon$ is the (Lorentz scalar) energy density, and $P = P(\varepsilon)$ is the equilibrium pressure defined via the equation of state. In astrophysical applications \cite{Rezzolla_Zanotti_book}, or in low-energy heavy-ion collisions \cite{An:2021wof}, one describes situations where the dynamics of the conserved baryon current, $J_B^\mu$, defined by $\nabla_\mu J^\mu_B=0$, becomes relevant and must be taken into account. In this work, for simplicity, we focus on relativistic hydrodynamic systems at zero baryon chemical potential (which provides a good approximation for heavy-ion collisions at very large center-of-mass energies) and, thus, fluid dynamics reduces to understanding the dynamics encoded in $T_{\mu\nu}$ only. Energy-momentum conservation gives four dynamical equations, which in the ideal case matches the number of independent dynamical variables and, thus, provides a closed system of equations. Strong hyperbolicity can be proven in different ways, and local well-posedness follows for the relativistic Euler equations \cite{Disconzi:2023rtt}. 

Just as in non-relativistic hydrodynamics, dissipative effects can be introduced through corrections in the form of tensor, vector, and scalar contributions, leading to a general energy-momentum tensor of the form $T^{\mu\nu} = T^{\mu\nu}_0 + \Pi^{\mu\nu}$. These dissipative effects are typically characterized by transport coefficients such as the shear viscosity $\eta$, bulk viscosity $\zeta$, and conductivity  $\sigma$, which quantify the response of the fluid to gradients in velocity, expansion, and temperature or chemical potential, respectively. The structure and the dynamics of the dissipative term $\Pi^{\mu\nu}$ depend on the specific formulation of relativistic fluid dynamics that is considered. Broadly speaking, modern approaches to relativistic viscous hydrodynamics fall into two categories: derivative-expansion formulations and the so-called second-order approaches. The most well-known examples of derivative-expansion formulations, corresponding to an attempt to generalize ordinary viscous fluid dynamics to the relativistic domain, were developed by Eckart \cite{EckartViscous} and Landau–Lifshitz \cite{LandauLifshitzFluids}. In this case, $\Pi^{\mu\nu}$ is defined by a \emph{constitutive relation} containing only first-order derivatives of the fluid dynamic variables that appear in ideal hydrodynamics, and these two formulations correspond to different ways to parametrize the viscous corrections (i.e. the hydrodynamic frame, see below). The equations of motion become a second-order system of nonlinear partial differential equations for the original hydrodynamic variables, which nicely reduce to standard Navier-Stokes theory in the appropriate non-relativistic regime \cite{Rezzolla_Zanotti_book}. However, despite their natural motivation and simplicity, the Eckart and Landau-Lifshitz formulations suffer from acausality and display unphysical instabilities for disturbances around equilibrium states \cite{Hiscock:1985zz}. 

The issues found in the relativistic Navier-Stokes formulations of Eckart and Landau and Lifshitz originally motivated a different approach in which $\Pi^{\mu\nu}$ is treated as a \emph{new set of dynamical variables}, which as such requires new equations of motion to close the dynamics. The prototypical example of such approach is Israel–Stewart (IS) theory \cite{Israel:1979wp}, where the new (relaxation-like) equations of motion for the dissipative fluxes are obtained by either expanding a suitably-defined non-equilibrium entropy density current to \emph{second order} in deviations from equilibrium, or via a truncation of the method of moments in relativistic kinetic theory \cite{Israel:1979wp, Denicol:2012cn}. For a review, see \cite{Rocha:2023ilf}. Modern variations of IS theory, such as \cite{Baier:2007ix,Denicol:2012cn}, are widely used in hydrodynamic simulations of the quark-gluon plasma formed in ultrarelativistic heavy-ion collisions \cite{Romatschke:2017ejr}. Second-order theories of Israel-Stewart-type also naturally appear when describing the dynamics of non-conserved currents in astrophysical systems \cite{Gavassino:2020kwo}, which has been particularly important to pin down possible effects coming from bulk viscosity due to weak interactions in neutron star mergers \cite{Alford:2017rxf,Most:2021zvc,Celora:2022nbp,Camelio:2022fds,Camelio:2022ljs,Most:2022yhe,Chabanov:2023blf,Chabanov:2023blf,Gavassino:2023xkt,Ripley:2023qxo,Yang:2023ogo,Yang:2025yoo}.

Second-order approaches indeed possess better physical and mathematical properties than the Eckart and Landau-Lifshitz formulations of viscous hydrodynamics. In the linear regime around equilibrium, causality and stability hold for judicious choices of parameters (defined by the equation of state and transport coefficients) \cite{Hiscock_Lindblom_stability_1983}, and for theories that stem from an information current principle \cite{Gavassino:2021cli,Gavassino:2021kjm}, strong hyperbolicity also holds near equilibrium \cite{Gavassino:2023odx}. In the nonlinear regime, only recently have general results concerning causality become available. Rigorous results establishing the conditions that ensure causality and strong hyperbolicity were found in the bulk-viscous case (no shear or diffusion) in \cite{Bemfica:2019cop} and in general-relativistic magnetohydrodynamic models of weakly collisional plasmas surrounding supermassive black holes \cite{Cordeiro:2023ljz}. General nonlinear causality conditions for second-order theories of diffusion were derived in \cite{Cordeiro:2025mtg} and, in Ref.\ \cite{Bemfica:2020xym}, for theories involving shear and bulk viscosity. The precise conditions under which causality, strong hyperbolicity, and local well-posedness hold in the nonlinear regime of second-order formulations, including all dissipative channels (shear, bulk, diffusion/heat), remain unknown. In particular, this becomes even more challenging when considering all the possible contributions to dissipative quantities that appear in second-order theories, see for instance the \emph{many} terms coming from the Knudsen and inverse Reynolds expansion discussed in \cite{Denicol:2012cn}. Finally, global well-posedness statements are limited to the case where only bulk viscosity is included\footnote{It was shown in \cite{Disconzi:2020ijk} that there exists a class of smooth initial data defined by localized perturbations of constant states for which the corresponding unique solutions to the Cauchy problem break down in finite time, or become unphysical.}, see Ref.\ \cite{Disconzi:2020ijk}. 

The so-called Bemfica-Disconzi-Noronha-Kovtun (BDNK) formulation \cite{Bemfica:2017wps, Kovtun:2019hdm, Bemfica:2019knx, Hoult:2020eho,Bemfica:2020zjp} of causal and stable hydrodynamics may provide a more manageable and simpler framework to determine how relativity affects dissipation, at least in situations near equilibrium. BDNK is the most general theory of viscous relativistic hydrodynamics with shear, bulk, and diffusion that includes first-order derivatives of the hydrodynamic variables (e.g. temperature $T$ and flow velocity $u^\mu$) in the constitutive relations while remaining causal, strongly hyperbolic, and stable\footnote{See also Refs.\ \cite{Disconzi2013,FreistuhlerTemple2014,CzubakDisconzi2016JMP,FreistuhlerTemple2017,FreistuhlerTemple2018,Freistuhler2020} for related mathematical work involving causality and hyperbolicity of first-order formulations.}. Local well-posedness has been proven in \cite{Bemfica:2019hok,Bemfica:2020gcl,Bemfica:2020zjp,Disconzi-Shao-2024} and global well-posedness for small initial data in Minkowski spacetime (though still in the full nonlinear regime) was established in \cite{Sroczinski,Freistuhler-Sroczinski-2025} at zero baryon density. The latter implies that the BDNK equations do exactly what one would expect a theory of relativistic viscous hydrodynamics should do: initial data describing small deviations from equilibrium evolve, in a nonlinear fashion, towards the equilibrium state in a way that is compatible with the fundamental principles of relativity. Such properties have not been established for any other relativistic viscous hydrodynamics framework. 

In BDNK, terms that formally involve first-order time derivatives in the constitutive relations are present, even in the local rest frame of the fluid. Such terms illustrate the freedom in determining the so-called hydrodynamic frame\footnote{This encodes our choice for the definition of what constitutes temperature, chemical potential, and flow velocity out of equilibrium, see \cite{Israel:1979wp,Bhattacharya:2011tra}.}, which emerges when one truncates the derivative expansion \cite{Kovtun:2012rj}. The additional transport parameters in BDNK theory, which are not just the usual $\eta$, $\zeta$, and $\sigma$, parameterize the choice of hydrodynamic frame, and in practice, simple choices involving a small number of parameters are possible without spoiling causality and stability, see \cite{Bemfica:2020zjp} and \cite{Abboud:2023hos}.

As mentioned above, BDNK hydrodynamics currently stands out as the only viscous formulation of relativistic fluid dynamics, including all dissipative effects, where causality, stability, and strong hyperbolicity properties are rigorously understood. An appealing feature of the BDNK framework is that causality can be ensured trivially by an appropriate choice of hydrodynamic frame parameters. This stands in sharp contrast to Israel–Stewart-type theories, where causality and stability must be verified through nontrivial inequalities that depend on the evolution of the system at each time step \cite{Bemfica:2020xym,Plumberg:2021bme}. This intrinsic simplicity makes BDNK not only mathematically appealing but also computationally convenient for numerical implementations. Furthermore, this makes it attractive for applications in both high-energy nuclear physics in the context of the hydrodynamic evolution of the quark-gluon plasma, and also in astrophysics (for instance, in the case of linearized perturbations, see \cite{Redondo-Yuste:2024vdb,Caballero:2025omv,Redondo-Yuste:2025ktt}). In fact, the equations of motion of BDNK hydrodynamics have been already solved numerically in many recent simulation studies, see \cite{Pandya_2021,Pandya_2022a,Pandya_2022b,Bantilan:2022ech,Bea:2023rru,Bea:2025eov,Bhambure:2024axa,Fantini:2025gnm,Keeble:2025bkc,Shum:2025jnl}. 

In this work, we introduce a new first-order flux-conservative formulation of BDNK relativistic viscous hydrodynamics, valid for both conformal and nonconformal fluids\footnote{Our approach can be readily extended to include baryon current effects. This is left for a future publication.}. Our focus here is on 1+1 simulations of  conformal BDNK hydrodynamics \cite{Bemfica:2017wps}, and we perform numerical simulations for two classes of initial conditions: smooth Gaussian energy profiles and semi-discontinuous stair-step configurations. To probe the regime of validity of BDNK, we systematically vary causal hydrodynamic frames, parameterized by transport parameters $\tau_\ep, \tau_\mathcal{Q}$ \cite{Bemfica:2020zjp}, and monitor the magnitude of Knudsen number-like quantities. Comparisons across hydrodynamic frames give a practical measure of \emph{hydrodynamic frame robustness} \cite{Bea:2023rru}, which provides a simple, practical way to estimate the regime of validity of the first-order BDNK truncation without having to resort to higher order terms in the derivative expansion. In fact, consistent with the independent study of frame robustness and shockwave profiles in first-order relativistic hydrodynamics performed in Ref.\ \cite{Bea:2023rru}, our results support frame-robust dynamics within the effective field theory regime.

Using this framework, we demonstrate that our flux-conservative formulation produces stable solutions free of spurious oscillations in the small-Knudsen regime. For Gaussian initial conditions at low viscosities and sufficiently large background energy density, the simulations recover the expected second-order numerical convergence, whereas for large viscosities and low background energy density, convergence degrades and nontrivial behavior in the Knudsen numbers emerges, matching the results indicated in \cite{Teaney:2024}. We show that these issues can be easily mitigated by simply raising the background energy density, restoring robust convergence and hydrodynamic frame robustness. Across all cases, velocity fields tend to converge more rapidly than energy density, while components of the energy–momentum tensor often display even higher effective convergence due to their dependence on gradients.

The existence and dynamical formation of shocks and other singularities in relativistic viscous fluids is a long-standing problem. Studies of late-time, steady-state solutions in Israel-Stewart theory --- where dissipation is the dominant effect --- have shown that discontinuous shock formation is precluded, leading instead to smooth profiles \cite{OLSON1990331}. Furthermore, causality and stability constraints impose strict bounds on shock dynamics \cite{GEROCH1991394}. These results, however, provide only a partial understanding, as they do not consider self-similar or fully dynamical solutions. This picture has been recently completed by an analysis of the early-time regime, where nonlinear modes dominate over dissipation. As shown in Ref. \cite{2025arXiv250804717B}, finite-time gradient blow-up and shock formation are indeed allowed in Israel-Stewart theory during early stages, a result also supported by numerical simulations. (See also \cite{Disconzi:2020ijk} for related work on finite-time breakdown of solutions in 3+1 dimensions). 
These problems have become more pressing, and interesting, in recent years given our new understanding of the nonlinear properties of Israel-Stewart and BDNK formulations. In particular, as we show here, BDNK can be expressed in a simple flux-conservative form, which is particularly well-suited for the investigation of shock formation.

In this paper, we also take the opportunity to initiate a study of how our flux-conservative formulation may be used to investigate shocks \cite{Freistuhler2021} and shock formation \cite{Pandya_2021} in BDNK theory. For sufficiently large viscosities/Knudsen numbers, frame robustness does not hold, and Gaussian initial data give rise to oscillations, while the stair-step initial conditions exhibit lower-order convergence ($p \sim 1$) consistent with the eventual formation of shock-like structures in a nearly frame-robust manner. These results support the idea that shocks can \emph{dynamically} develop within BDNK fluids, complementing recent findings in other spacetimes \cite{Keeble:2025bkc}. Interestingly, we also observe that there is initial data for which viscous effects in BDNK prevent the formation of sharp gradients in a frame robust manner. This is particularly notable given that BDNK is a hyperbolic system of equations, where one would not generally expect diffusive-like smoothing behavior characteristic of parabolic PDEs. While the precise dynamical characterization of viscous shock formation in BDNK theory remains an open problem, the present analysis establishes a framework for systematically studying their emergence, with numerical convergence testing serving as a potential diagnostic for simulation reliability and hydrodynamic frame robustness. Alongside this, the fact that viscosity in this context acts as an effective regularizing mechanism underscores how viscous dissipation can coexist and interact nontrivially with nonlinear hyperbolic wave propagation in relativistic fluid dynamics.

This paper is organized as follows. In Section \ref{section:BDNKoverview}, we present an overview of BDNK at zero chemical potential and introduce the novel flux conservative formulation that reduces the order of the BDNK equations while casting them into a flux-conservative form.  The numerical simulations in this work focus on the conformal case, and details of this implementation are also found in the same section as above. Section \ref{section:numericalimplementation} discusses how to implement our approach for 1+1 Cartesian dynamics, along with the form of the flux conservative equations and the numerical methods employed. Our numerical results are found in Section \ref{section:Results}, along with our choices for the initial conditions we use and a discussion about Knudsen numbers. The conclusions and future directions can be found in Section \ref{section:Conclusion}.  Further details about the evolution of each initial condition are given in Appendix \ref{appendix:gaussian}. Following this, in Appendix \ref{appendix:Euler} we  outline how we numerically solved the equations of motion of relativistic ideal fluid dynamics, which we used to compare to our BDNK results. In Appendix \ref{appendix:burgers}, we present a pedagogical discussion of how damping can preclude shock formation for small data in the damped hyperbolic Burgers equation in 1+1 dimensions. This is particularly interesting as it provides a simple, yet nontrivial example of how dissipation affects shock formation in dissipative hyperbolic equations, and how that depends on the initial data. A detailed overview of our convergence testing procedure is covered in Appendix \ref{appendix:convergence}. Finally, in Appendix \ref{appendix:pandya_agreement} we demonstrate that our simulations accurately reproduce well-known results previously published in the literature \cite{Pandya_2021}.

\subsection*{Note added:} While this manuscript was being finalized, a different 3+1 flux conservative formulation of BDNK hydrodynamics was presented in Ref.\ \cite{Shum:2025jnl} and applied to perform numerical simulations
of spherically symmetric neutron stars in the Cowling approximation. This is, however, very different than the focus of the present work.  

\subsection*{Notation} 
We use a mostly plus metric signature, 4-dimensional Minkowski spacetime in Cartesian coordinates, and natural units $\hbar=c=k_B=1$.

\section{Flux conservative BDNK hydrodynamics at zero chemical potential} 
\label{section:BDNKoverview}

A key difference of BDNK hydrodynamics with respect to the relativistic Navier-Stokes theories due to Eckart \cite{EckartViscous}, and  Landau and Lifshitz \cite{LandauLifshitzFluids}, is the systematic inclusion of all possible out-of-equilibrium corrections to the energy-momentum tensor 
\cite{Bemfica:2017wps, Kovtun:2019hdm, Bemfica:2019knx, Hoult:2020eho, Bemfica:2020zjp}:
\be
T_{\mu\nu} = \left(\varepsilon +\cA \right) u_\mu u_\nu + (P+\Pi)\Delta_{\mu\nu}
+\pi_{\mu\nu}
+u_\mu \mathcal{Q}_\nu + u_\nu \mathcal{Q}_\mu\,,
\ee
where  
$\mathcal{A}$ is the out-of-equilibrium correction to the energy density, $\Pi$ is the bulk-viscous pressure,  $\mathcal{Q}_\mu$ is the energy diffusion 4-vector (defined such that $\mathcal{Q}_\mu u^\mu=0$),  
and $\pi_{\mu\nu}$ is the symmetric and traceless shear stress tensor, which obeys $  u_\mu \pi^{\mu\nu}=0$.

We focus on the case of vanishing chemical potential $\mu=0$.
In the spirit of an effective field theory framed in the context of a derivative expansion, in BDNK, dissipative corrections are defined by constitutive relations given by the most general expressions involving first-order derivatives of the hydrodynamic fields (here, $\varepsilon$ and $u^\mu$), which we write as:
\begin{subequations}
\begin{align}
   & \cA = \tau_{\ep 1}\,  u^\mu \nabla_\mu \varepsilon + \tau_{\ep 2}\,(\varepsilon+P) \nabla_\mu u^\mu ,
&
\\
&
    \Pi =  \tau_{P 1}\,
    u^\lambda \nabla_\lambda \ep + \tau_{P 2}\,(\ep + P) \nabla_\lambda u^\lambda ,
& 
\\
&
    \mathcal{Q}_\nu = \tau_{\mathcal{Q} 1} \,(\varepsilon+P)u^\lambda \nabla_\lambda u_\nu + \tau_{\mathcal{Q} 2}\, \Delta_{\nu}^\lambda \nabla_\lambda P,
\end{align}
\end{subequations}
where the shear-stress tensor is given by $\pi_{\mu\nu}=-2\eta\,\sigma_{\mu\nu}$, $\eta$ is the shear viscosity and 
$\sigma_{\mu\nu}=\Delta_{\mu\nu}^{\alpha\beta}\nabla_\alpha u_\beta$ is the shear tensor, written in terms of the the rank-4 projector $\Delta_{\mu\nu}^{\alpha\beta} = \left(\Delta^\alpha_\mu \Delta^\beta_\nu + \Delta^\alpha_\nu \Delta^\beta_\mu\right)/2 - \Delta_{\mu\nu}\Delta^{\alpha\beta}/3$. 

The coefficients $\tau_{(\ldots)}$ encode true first-order transport coefficients, such as $\zeta$ and $\sigma$, as well as parameters responsible for regulating the ultraviolet behavior of the theory, defining a hydrodynamic frame --- that is,  a local-equilibrium state with respect to which the derivative expansion is defined. Changing the hydrodynamic frame amounts to shifting the definition of the local temperature and flow velocity by terms of order $\mathcal{O}(\partial)$, and leads to mere redefinitions of these coefficients, as long as terms  $\mathcal{O}(\partial^2)$ and higher are properly discarded \cite{Kovtun:2019hdm}. 
For a more in-depth discussion of hydrodynamic frames, we refer the reader to \cite{Kovtun:2019hdm,Bemfica:2020zjp}. 

Not all hydrodynamic frames lead to causal and stable behavior in BDNK theory. 
The Landau frame, for instance, which treats time and spatial derivatives on a different footing in the local rest frame,  is known to lead to acausal and unstable behavior \cite{Hiscock_Lindblom_instability_1985}.
The allowed ranges for the BDNK parameters, dictated by causality and linear stability, have been derived in \cite{Bemfica:2017wps, Kovtun:2019hdm, Bemfica:2019knx,Hoult:2020eho,Bemfica:2020zjp}. For the proofs concerning strong hyperbolicity and local well-posedness in the full nonlinear regime, including all dissipative effects, see \cite{Bemfica:2020zjp}. 

Thermodynamic consistency considerations imply that $\tau_{\mathcal{Q} 1} = \tau_{\mathcal{Q} 2} = \tau_\mathcal{Q}$ \cite{Kovtun:2019hdm} (so that not only $\mathcal{A}$ and $\Pi$, but also $\mathcal{Q}$,  vanish in the equilibrium state). 
Further choosing $\tau_{\ep 1} = \tau_{\ep 2} = \tau_\ep$ and $\tau_{P 1} = \tau_{P 2} = \tau_P$ makes $\mathcal{A}$, $\Pi$ and $\mathcal{Q}$ vanish for solutions to the equations of motion of ideal hydrodynamics. 
Since deviations from ideal hydrodynamics are of order $\mathcal{O}(\partial)$ themselves, this choice makes $\mathcal{A}$, $\Pi$ and $\mathcal{Q}$ 
effectively  $\mathcal{O}(\partial^2)$ when evaluated on-shell --- that is, evaluated on solutions to the equations of motion.
In this work, we follow \cite{Bemfica:2020zjp} and define $\tau_P = c_s^2 \tau_\varepsilon$, where $c_s = \sqrt{dP/d\varepsilon}$ is the equilibrium speed of sound. In this case, the out-of-equilibrium correction to the energy density can be written as
\begin{subequations}
\be
\cA = \tau_\ep \left [ u^\mu \nabla_\mu \varepsilon + (\varepsilon+P) \nabla_\mu u^\mu \right ],
\ee
while the bulk viscous scalar is
\be
\Pi =  -\zeta \nabla_\lambda u^\lambda + \tau_\varepsilon c_s^2 \left [
u^\lambda \nabla_\lambda \ep + (\ep + P) \nabla_\lambda u^\lambda 
\right ],
\ee
and 
\be
\mathcal{Q}_\nu = \tau_\mathcal{Q} \left[(\varepsilon+P)u^\lambda \nabla_\lambda u_\nu + \Delta_{\nu}^\lambda \nabla_\lambda P\right].
\ee
\end{subequations}
Above, $\eta$ and $\zeta$ are the (frame invariant) shear and bulk viscosity coefficients (which may depend on $\varepsilon$) that control the dissipation of sound waves, as in Navier-Stokes theory \cite{Kovtun:2019hdm}. The BDNK parameters $\{ \tau_\ep, \tau_\mathcal{Q} \}$ are functions of $\varepsilon$ that, in practice, define the temperature and flow velocity out of equilibrium, setting the hydrodynamic frame.

The equations of motion that govern the fluid stem from energy-momentum conservation, $\nabla_\mu T^{\mu\nu}=0$, and are obtained by taking the parallel and orthogonal components with respect to the flow,
\bml
\bea 
\label{EOMenergy}
&&u^\alpha \nabla_\alpha \mathcal{A} + \nabla_\alpha \mathcal{Q}^\alpha + \left(\varepsilon+P+\mathcal{A}+\Pi\right)\nabla_\alpha u^\alpha + u^\alpha \nabla_\alpha \varepsilon - 2\eta \sigma^{\alpha\beta}\sigma_{\alpha\beta} + \mathcal{Q}_\alpha u^\beta \nabla_\beta u^\alpha = 0 ,\\ \label{EOMvector}
&& u^\beta \nabla_\beta \mathcal{Q}^\alpha + \Delta^{\alpha\beta}\nabla_\beta \Pi - 2\eta \nabla_\beta \sigma^{\alpha\beta} + (\varepsilon+P+\mathcal{A}+\Pi)u^\beta \nabla_\beta u^\alpha + c_s^2 \Delta^{\alpha\beta}\nabla_\beta \varepsilon - 2\sigma^{\alpha\beta}\nabla_\beta \eta\nonumber \\  &+& 2\eta u^\alpha \sigma^{\mu\nu} \sigma_{\mu\nu} +\mathcal{Q}^\beta \nabla_\beta u^\alpha-  u^\alpha \mathcal{Q}^\beta u^\mu \nabla_\mu u_\beta + \mathcal{Q}^\alpha \nabla_\beta u^\beta=0. 
\eea
\eml

\subsection{Flux conservative formulation} \label{section:flux-conservative-formulation}

Let us now consider the vector $\beta_\mu = u_\mu /T$. From its definition, we find that in general
\be
\begin{split}
\nabla_\mu \beta_\nu + \nabla_\nu \beta_\mu  = & -\frac{u_\mu}{T}\left(u^\lambda \nabla_\lambda u_\nu +\frac{\Delta_\nu^\lambda \nabla_\lambda T}{T}\right) - \frac{u_\nu}{T}\left(u^\lambda \nabla_\lambda u_\mu + \frac{\Delta_\mu^\lambda \nabla_\lambda T}{T}\right) + 2\frac{u_\mu u_\nu}{T^2}(u^\lambda \nabla_\lambda T) 
\\
& + 2\frac{\sigma_{\mu\nu}}{T} + \frac{2}{3T}\Delta_{\mu\nu}(\nabla_\lambda u^\lambda).
\end{split} 
\ee
In equilibrium, $\beta_\mu$ is a Killing vector \cite{Israel:1979wp}, i.e.
\be
\nabla_\mu \beta_\nu + \nabla_\nu \beta_\mu=0,
\ee
which implies that in equilibrium 
\be
u^\mu \nabla_\mu T = 0, \qquad \nabla_\mu u^\mu = 0, \qquad \sigma_{\mu\nu}=0, \qquad u^\lambda \nabla_\lambda u_\mu + \frac{\Delta_\mu^\lambda \nabla_\lambda T}{T}=0.
\ee
The vector $\beta_\mu$, as defined above, proves to be useful as it offers a way to nicely express the viscous corrections of the BDNK energy-momentum tensor in terms of deviations from the Killing condition:
\be \label{BDNK}
T^{\mu\nu} = T_0^{\mu\nu} + H^{\mu\nu\alpha\rho}\nabla_\alpha \beta_\rho
\ee
where the ``viscous susceptibility tensor'' is defined as
\bea
H^{\mu\nu\alpha\beta} &=& \frac{\tau_\varepsilon(\varepsilon+P)T}{c_s^2}\left(u^\mu u^\nu + c_s^2 \Delta^{\mu\nu}\right)\left(u^\alpha u^\beta + c_s^2 \Delta^{\alpha\beta}\right)  - \zeta T \Delta^{\mu\nu}\Delta^{\alpha\beta} - 2\eta T \Delta^{\mu\nu\alpha\beta} \nonumber \\ &+& \tau_\mathcal{Q} T (\varepsilon+P) \left(u^\mu u^\beta \Delta^{\alpha\nu}+u^\nu u^\beta \Delta^{\mu\alpha}+u^\mu u^\alpha \Delta^{\beta\nu}+u^\nu u^\alpha \Delta^{\mu\beta}\right).
\label{defineBDNKnew}
\eea
Both $T^{\mu\nu}_0$ and $H^{\mu\nu\alpha\beta}$ depend solely on $\beta_\mu$, with $\beta = \sqrt{-\beta_\mu \beta^\mu} = 1/T$. The tensor $H^{\mu\nu\alpha\beta}$ has dimension $T^4$ and is symmetric under the exchanges $\mu \leftrightarrow \nu$, $\alpha \leftrightarrow \beta$, and $(\mu\nu) \leftrightarrow (\alpha\beta)$. The physical  content of the theory is fully specified by the equation of state $P(\varepsilon)$, along with the shear viscosity $\eta$ and bulk viscosity $\zeta$. The BDNK parameters $\tau_\varepsilon$ and $\tau_\mathcal{Q}$ are chosen to ensure causality and stability. Using the conservation law $\nabla_\mu T^{\mu\nu} = 0$, it follows from Eq.\ \eqref{defineBDNKnew} that the principal part of the BDNK equations is governed by $H^{\mu\nu\alpha\beta}$, and, thus, the conditions for causality are directly derived from the structure of $H^{\mu\nu\alpha\beta}$. Using the results from \cite{Bemfica:2020zjp}, it follows that for causal and stable frames, the matrix $H^{\mu00\nu}$ always has an inverse (see the proof in Sec.\ \ref{Sec:Causality}), a fact that will be used below in our flux conservative formulation.

Casting the BDNK equations in this form demonstrates that dissipation occurs only if $\beta_\mu$ is not a Killing vector. In our numerical simulations, we choose to work with an alternative, yet closely related field,
\begin{equation}
    C_\mu = T u_\mu = T^2 \beta_\mu.
\end{equation}
This choice ensures the solution remains well-behaved in regions where $T$ may become very small, such as in the boundaries of the simulation domain. It is important to note that the normalization condition $u_\mu u^\mu = -1$ leads to the relation $T = \sqrt{-C_\mu C^\mu}$, which means we can determine the temperature at any time step of the simulation, as $C_\mu$ will be an evolved quantity in the BDNK solution algorithm, which will be defined shortly. 

In order to express our PDEs in first-order form, we must also evolve the derivatives of $C_\mu$, which prompts us to define
\bea \label{EOM_Cm}
X_{\mu\nu} = \partial_\mu C_\nu,
\eea
an object that is not, in general, symmetric. We can now express derivatives of $\beta_\mu$ in terms of $X_{\mu\nu}$, resulting in
\be
\partial_\alpha \beta_\rho = \frac{X_{\alpha\rho}}{T^2} + \frac{2}{T^2}u_\rho u^\lambda X_{\alpha\lambda}=\frac{1}{T^2}\left (\delta^\lambda_\rho+2u_\rho u^\lambda\right )X_{\alpha\lambda}.
\ee
This allows us to rewrite \eqref{BDNK} as
\bea
T^{\mu\nu} &=& T^{\mu\nu}_0 + \frac{1}{T^2}H^{\mu\nu \alpha\rho}\left (\delta_\rho^\lambda+2u_\rho u^\lambda\right )X_{\alpha\lambda}- \frac{1}{T^2}H^{\mu\nu\alpha\rho} \Gamma_{\alpha\rho}^\lambda C_\lambda,
\label{defineTnew}
\eea
where $\Gamma^\lambda_{\alpha\rho}$ is the standard Christoffel symbol.
We can now write the equations of motion in first-order form using energy-momentum conservation,
$\nabla_\mu T^{\mu\nu} = 0$, together with  \eqref{EOM_Cm}. However, these 8 dynamical equations are insufficient to evolve the full set of 20 independent variables ${C_\mu, X_{\mu\nu}}$. To obtain a complete set of evolution equations, we use the definition of $X_{\mu\nu}$ in \eqref{EOM_Cm} together with the identity $\partial_\alpha \partial_\beta C_\nu = \partial_\beta \partial_\alpha C_\nu$ to obtain  
\bea \label{defineEqXij}
\partial_0 X_{i\nu} - \partial_i X_{0\nu}=0.
\eea
Although we have reduced the second-order partial differential equations (PDEs) for the variables $C_\mu$ to a first-order system of PDEs for the set $\{C_\nu, X_{\mu\nu}\}$, our main focus is to obtain a flux-conservative set of equations and a natural candidate for this is to choose $T^{0\nu}$ as variables in place of $X_{0\nu}$.
From \eqref{defineTnew}, we can find $T^{0\nu}$ in terms of variables $X_{0\alpha}$, $C_\nu$, and $X_{i\nu}$. Thus, we need to find the inverse relation $X_{0\nu} = X_{0\nu}(T^{0\alpha})$ in order to promote $T^{0\nu}$ as dynamical variables instead of $X_{0\nu}$.
The advantage of this approach lies in the fact that we may split $\nabla_\nu T^{\mu\nu}=\partial_\nu T^{\mu\nu}+\Gamma^\mu_{\nu\lambda}T^{\lambda \nu}+\Gamma^\nu_{\nu\lambda}T^{\mu\lambda} = 0$ into the flux-conservative forms 
\bml
\label{Conservation_Eq}
\bea
\partial_0 T^{00} + \partial_i T^{0i} &=& -\Gamma^0_{\mu\lambda}T^{\lambda \mu}-\Gamma^\mu_{\mu\alpha}T^{0\lambda},\\
\partial_0 T^{0i} + \partial_k T^{ik} &=& -\Gamma^i_{\mu\lambda}T^{\lambda \mu}-\Gamma^\mu_{\mu\lambda}T^{i\lambda}, 
\eea
\eml
where $T^{ik}$ depends on all the dynamical variables. As shown in Sec.\ \ref{Sec:Causality}, the matrix $H^{0\nu0\rho}=H^{\nu00\rho}$ is invertible when causality holds. Therefore, assuming causal transport coefficients that obey $c_a^2 \in [0,1]\subset\mathbb{R}$, with $c_a$ defined in \eqref{Characteristics_general}, we may define the invertible matrix \begin{align*}
M^{\mu\beta} \equiv H^{\mu 00\beta}&=A u^\mu u^\beta + B \left (u^\mu  \Delta^{0\beta}
+ \Delta^{\mu 0}u^\beta\right )   + C \Delta^{0\mu }\Delta^{0\beta}+ D\Delta^{\mu \beta}, 
\end{align*}
where
\begin{align*}
A&=\frac{u_0^2(\varepsilon+P)\tau_\varepsilon T}{c_s^2}+ (\varepsilon+P)\tau_\mathcal{Q} T \Delta^{00},\\
B&=u^0 (\varepsilon+P)(\tau_\varepsilon+\tau_\mathcal{Q})T,\\
C&= c_s^2 (\varepsilon+P)\tau_\varepsilon T- \zeta T - \frac{\eta T}{3},\\
D&=(\varepsilon+P)\tau_\mathcal{Q} T u_0^2 - \eta T \Delta^{00}.   
\end{align*}
The inverse $M^{-1}_{\sigma\mu}$, which obeys $M^{-1}_{\sigma\mu} M^{\mu\beta} = \delta_\sigma^\beta$, is 
\begin{align*}
M^{-1}_{\sigma\mu}&=\frac{C \Delta^{00}+D}{E}u_\sigma u_\mu + \frac{B}{E}(u_\sigma \Delta^0_\mu+\Delta^0_\sigma u_\mu )+\frac{B^2-AC}{DE}\Delta^0_\sigma\Delta^0_\mu+\frac{1}{D}\Delta_{\sigma\mu}, 
\end{align*}
where $E=A D+(A C-B^2) \Delta^{00}$. This inverse, together with the identity
\[
(\delta^\sigma_\beta + 2 u_\beta u^\sigma)(\delta_\sigma^\lambda + 2 u_\sigma u^\lambda) = \delta^\lambda_\mu,
\]
allows us to express $X_{0\nu} = X_{0\nu}(T^{0\mu})$ via \eqref{defineTnew}. We first set $\mu=0$ in \eqref{defineTnew}, then multiply by $T^2(\delta^\sigma_\beta + 2 u_\beta u^\sigma) M^{-1}_{\sigma\rho}$ and isolate $X_{0\beta}$ to obtain
\bea
X_{0\beta} &=& (\delta^\sigma_\beta + 2 u_\beta u^\sigma) M^{-1}_{\sigma\nu} \left[ T^2 (T^{0\nu} - T^{0\nu}_0)
- H^{0\nu i\rho} (\delta_\rho^\lambda + 2 u_\rho u^\lambda) X_{i\lambda} + H^{0\nu\alpha\rho} \Gamma_{\alpha\rho}^\lambda C_\lambda \right].
\label{X_Inverse}
\eea

Eqs.\ \eqref{Conservation_Eq} together with the auxiliary equations \eqref{defineEqXij} form a complete set of first-order and flux-conservative equations of motion for the set $\{T^{0\nu},C_\nu,X_{i\alpha}\}$, while also having a source term without any derivatives of the evolved fields. One may express the system of equations as follows: 
\be \label{EOM_vec}
\partial_0 \m{ T^{0\nu} \\ C_0 \\ C_i \\ X_{ij} \\ X_{i0} }
+ \partial_k \m{  T^{k\nu} \\ 0 \\ 0_i \\ -\delta^k_i X_{0j} \\ - \delta^k_i X_{00} }
= \m{ - \Gamma ^\mu_{\mu\lambda} T^{\lambda \nu} - \Gamma^\nu_{\mu\lambda} T^{\mu\lambda} \\ X_{00} \\ X_{0i} \\ 0_{ij} \\ 0_{i} \\},
\ee
where $0_i$ and $0_{ij}$ are zero column vectors of size $d\times 1$ and $d^2\times 1$, respectively, with $d$ being the number of space coordinates. 
This formulation in principle may be used in a general spacetime and in different spacetime dimensions, although here in our work we focus only on the 4-dimensional Minkowski case in Cartesian coordinates. Note that the source term depends only on the dynamical fields themselves, not on their derivatives.

We remark that, to completely determine both the flux and the source, it is necessary at each iteration in the simulation to obtain $X_{0\mu}$, which we do through \eqref{X_Inverse}. This can always be done because we have already evolved $T^{0\nu}, C_\mu, X_{ij}$ and $X_{i0}$ at each time step and the required matrix $M^{\nu\rho}$ is guaranteed to have an inverse in hydrodynamic frames that are causal, as shown in Sec.\ \ref{Sec:Causality}. 

We must point out an important situation that occurs in numerical simulations where the fields depend on more than one spatial coordinate. By introducing extra dynamical fields defined in terms of derivatives, such as $X_{\mu\nu}$, we inherit the constraints $ \partial_\mu X_{\nu\alpha} - \partial_\nu X_{\mu\alpha} = 0$. The auxiliary equations in \eqref{defineEqXij} dynamically evolve parts of these constraints. However, the constraints $ \partial_i X_{j\alpha} - \partial_j X_{i\alpha} = 0$ do not appear anywhere in the evolution equations. Mathematically, this is not a problem as long as these constraints are consistently set to zero in the initial data. However, one must take care to ensure that, numerically, these curl-like constraints for the new variables remain satisfied. One way to address this numerical issue is by means of Generalized Lagrange Multiplier (GLM) methods, which introduce extra dynamical ``cleaning'' fields. We summarize this procedure in Section \ref{section:CurlCleaning} for completeness. It is worth mentioning, however, that GLM methods are not necessary for our 1+1-dimensional simulations because, in this case, the extra constraints that are not dynamically evolved are not present; indeed, $\partial_1 X_{1\mu} - \partial_1 X_{1\mu} = 0$ is trivially satisfied as an identity.

Finally, we note that the energy-momentum tensor of BDNK has first-order derivatives of the hydrodynamic variables and, thus, the conservation laws imply the equations of motion that govern BDNK are second-order PDEs (note that the same occurs in Eckart or Landau-Lifshitz theories). In this section, we have shown how one can nicely recast the \emph{second-order} BDNK PDEs into a set of \emph{first-order} PDEs in an exactly flux-conservative form, giving us access to well-known numerical tools useful for such a set of PDEs, such as a straightforward implementation of the Kurganov-Tadmor algorithm, see Sec.~\ref{section:KT}. In fact, our equations are cast in conservation law form,
\bea \label{EOM_CL}
\partial_0 \mathbf q + \partial_k \mathbf F ^{k}[\mathbf q] = \mathbf S [ \mathbf q].
\eea 
where the state vector $\mathbf q$, the flux vector $\mathbf F^k$, and the source vector $\mathbf S$ are defined by \eqref{EOM_vec}, with the latter two being generally functionals of the state vector.  

\subsection{Causality conditions}
\label{Sec:Causality}

From the definition of the energy-momentum tensor in Eq.\ \eqref{BDNK}, the equations of motion that follow from its conservation law $\nabla_\nu T^{\mu\nu}=0$ yield the quasi-linear second-order PDEs
\be 
\label{Causality1}
H^{\mu\nu\alpha\rho}\partial_\nu\partial_\alpha \beta_\rho=B^{\mu}(\partial\beta,\partial^2g),
\ee
where $B^\mu$ contains all lower-order terms in derivatives of $\beta^\nu$. Since we are assuming a given  background metric $g_{\mu\nu}$, it does not contribute to the principal part here, which only contains terms of second-order derivatives. However, as shown in \cite{Bemfica:2020zjp}, when the gravitational field is considered dynamical and $\partial^2 g$ contributes to the principal part of \eqref{Causality1}, the characteristic roots are not affected, because the matter sector and the gravity sector split in such a way that this contribution gets canceled out in the calculation of the characteristics. For completeness, we recompute the analysis performed in \cite{Bemfica:2017wps,Bemfica:2019knx,Bemfica:2020zjp}, as different nomenclature has been used for the transport coefficients across these papers. Our present notation follows the relaxation time definitions of \cite{Bemfica:2020zjp}. 

Let $\beta_\nu$ be a solution of $\nabla_\nu T^{\mu\nu}=0$ for $T^{\mu\nu}$ given in  \eqref{BDNK}, with $\beta_\nu$ a timelike covector so that the equations for $T=1/\sqrt{-\beta_\nu \beta^\nu}$ and $u^i=\beta^i/\sqrt{-\beta_\nu \beta^\nu}$ directly follow. The causality constraints are determined by analyzing the roots of the characteristic equation $\det[\tensor{H}{^\mu^\nu^\alpha_\beta}\xi_\nu\xi_\alpha]=0$, where $\xi_\mu=\partial_\mu \phi$ is a covector orthogonal to the characteristic surface $\Sigma=\{\phi(\mathbf{x})=\text{constant}\}$ of the system. We omit further mathematical details here and refer the reader to \cite{Bemfica:2017wps,Bemfica:2019knx,Bemfica:2020zjp}. Nonlinear causal solutions are guaranteed if, and only if, all the characteristic roots $\xi_0=\xi_0(\bm{\xi})$ of the above characteristic determinant are real (hyperbolic equations) and define a non-timelike covector $\xi_\mu$ (timelike or null-like characteristic surface). By defining $a=u^\alpha\xi_\alpha$, $v^\mu=\Delta^{\mu\nu}\xi_\nu$, and $v^2=\Delta^{\mu\nu}\xi_\mu\xi_\nu$, we obtain that
\bea
\det[\tensor{H}{^\mu^\nu^\alpha_\beta} \xi_\nu \xi_\alpha]
&=&T^4\det[ F \delta^{\mu}_{\beta}+ V^\mu u_{\beta} +  W^\mu v_\beta],\nonumber\\
&=&T^4 F^2 \left [F^2+F(V^\mu u_\mu+W^\mu v_\mu)+(V^\mu u_\mu)(W^\nu v_\nu)-(V^\mu v_\mu)(W^\nu u_\nu)\right ]\nonumber\\
&=&-\frac{T^4\tau_\varepsilon\tau_\mathcal{Q}^3(\varepsilon+P)^4}{c_s^2}\prod_{a=1,+,-}[(u^\alpha\xi_\alpha)^2-c_a^2\Delta^{\mu\nu}\xi_\mu\xi_\nu]^{n_a},
\eea
where
\begin{align*}
F&=\tau_\mathcal{Q}\rho a^2-\eta v^2,\\
V^\mu&=(\tau_\mathcal{Q} \rho a^2-\eta v^2) u^{\mu}+\frac{\tau_\varepsilon\rho }{c_s^2}\left(u^\mu a + c_s^2 v^\mu \right)a+\tau_\mathcal{Q} \rho v^2 u^\mu +\tau_\mathcal{Q} \rho a v^\mu,\\
W^\mu&=\frac{\tau_\varepsilon\rho }{c_s^2}\left(u^\mu a + c_s^2 v^\mu \right)c_s^2- \zeta  v^\mu -\frac{\eta}{3}v^\mu +\tau_\mathcal{Q}  \rho a u^\mu,    
\end{align*}
$n_1=2$, $n_\pm=1$, and the characteristic velocities squared are
\bml
\label{Characteristics_general}
\begin{align}
c_1^2&=\frac{\eta}{\tau_{\mathcal{Q}}(\varepsilon+P)},\\
c_\pm^2&=\frac{\tau_\varepsilon \left[6 (\varepsilon+P)  c_s^2 \tau _Q+3 \zeta +4 \eta\right]\pm\sqrt{\Delta}}{6(\varepsilon+P)\tau_\varepsilon\tau_{\mathcal{Q}}},\\
\Delta&=\tau_\varepsilon (3 \zeta +4 \eta ) \left\{\tau_\varepsilon \left[12 (\varepsilon+P)  c_s^2 \tau_{\mathcal{Q}}+3 \zeta +4 \eta \right]+12 (\varepsilon+P)  c_s^2\tau_{\mathcal{Q}}^2\right\}.
\end{align}
\eml
Causality follows directly from condition $c_a^2\in[0,1]\subset \mathbb{R}$ together with $T,\tau_\varepsilon,\tau_\mathcal{Q},\varepsilon+P,c_s\ne 0$. The proof relies on the fact that the characteristic determinant $\det[\tensor{H}{^\mu^\nu^\alpha_\beta}\xi_\nu\xi_\alpha]$ factors into a product of second-degree polynomials in $\xi_0$ of the form $(u^\mu\xi_\mu)^2-c_a^2\Delta^{\mu\nu}\xi_\mu\xi_\nu$. For each $c_a$ in \eqref{Characteristics_general}, we obtain the characteristic roots $\xi_0=\xi_0(\bm{\xi})$ of the corresponding factored polynomial. We actually do not need to compute explicitly the roots once we are able to prove that they are real and the corresponding covector $\xi_\mu$ is non-timelike. We again omit the details and quote \cite{Bemfica:2020zjp}, where it has been shown that $\xi_0\in\mathbb{R}$ if $c_a^2\in[0,1]\subset \mathbb{R}$. We further note that all $c_a$ are real when $\varepsilon+P>0$, $\tau_\varepsilon>0$, $\tau_{\mathcal{Q}}>0$, $c_s^2>0$, and $\zeta,\eta\ge 0$, which ensures $\Delta\ge 0$. Furthermore, substituting $(u^\mu\xi_\mu)^2=c_a^2\Delta^{\mu\nu}\xi_\mu\xi_\nu$ into $\xi_\mu \xi^\mu$ we obtain $\xi_\alpha\xi^\alpha=-(u^\alpha\xi_\alpha)^2+\Delta^{\mu\nu}\xi_\mu\xi_\nu=(1-c_a^2)\Delta^{\mu\nu}\xi_\mu\xi_\nu$. Consequently, for each one of the roots $\xi_0=\xi_0(\bm{\xi})$, the corresponding covector $\xi_\mu$ is non-timelike ($\xi_\mu \xi^\mu\ge 0$) if, and only if, $c_a^2\in[0,1]$ since $\Delta^{\mu\nu}\xi_\mu\xi_\nu\ge 0$ for any real $\xi_\mu$. 

Given the causality constraints $c_a^2 \in [0,1]\subset\mathbb{R}$, where the $c_a$ are defined in \eqref{Characteristics_general} for the general barotropic fluid, the matrix $H^{\mu 00 \rho}$ is invertible. This can be seen by considering the timelike covector $\omega_\alpha = (1,0,0,0)$. Causality ensures that $\omega_\alpha$ cannot be a root of the characteristic polynomial, meaning $\det[\tensor{H}{^\mu^\nu^\alpha_\beta} \omega_\nu \omega_\alpha] \ne 0$. Since $\det[H^{\mu 00 \rho}]=\det[g^{\rho\beta}]\det[\tensor{H}{^\mu^\nu^\alpha_\beta}]$, it must also be non-zero. Therefore, the matrix $H^{\mu 00 \rho}$ is invertible under the causality constraints.

For the conformal case, let us compare our results with those in \cite{Bemfica:2017wps}. To do this, we define $\chi\equiv (\varepsilon+P)\tau_\varepsilon$ and $\lambda\equiv(\varepsilon+P)\tau_{\mathcal{Q}}$. In this case, $\zeta=0$, $c_s^2=1/3$, $P=\varepsilon/3$, and the characteristic velocities are simply
\begin{align}
\label{Characteristics_conformal}
c_1^2=\frac{\eta}{\lambda}\quad\text{and}\quad
c_\pm^2=\frac{\chi(\lambda+2\eta)\pm 2\sqrt{\eta  \chi  \left(\eta  \chi+\lambda  \chi +\lambda ^2 \right)}}{3\lambda\chi}.
\end{align}
Assuming $\eta>0$, the causality conditions $c_a^2\in[0,1]$ result in the inequalities
\begin{align}
\label{Causality_conformal}
\chi\ge 4\eta\quad\text{and}\quad\lambda\ge \frac{3\chi \eta}{\chi-\eta}.
\end{align}
It is worth mentioning that \eqref{Causality_conformal} guarantees $\lambda\ge \eta$, which implies $c_1^2\in(0,1)$.

\subsection{Equivalence between the new system of equations and BDNK} \label{section:proof}

The equations of motion in \eqref{EOM_vec} in principle could have more solutions than the corresponding BDNK equations, even in 1+1 dimensions. To ensure that solutions of \eqref{EOM_vec} are solutions of BDNK we need 
\be
S_{\mu\nu} = X_{\mu\nu} - \partial_\mu C_\nu
\ee
to vanish in the initial condition, and it must remain zero throughout the evolution. In practice, besides tailoring the initial data such that $S_{\mu\nu}$ vanishes, we need our evolution equations \eqref{EOM_vec} to imply that 
\be
\partial_t S_{\mu\nu} = 0.
\ee
From the equations of motion \eqref{EOM_vec}, we have $\partial_0 C_0 = X_{00}$ and $\partial_0 C_i = X_{0i}$ so we see that indeed $\partial_t S_{0\nu}=0$. Thus, if we set $S_{00}$ and $S_{0i}$ to zero at the initial condition, they will remain zero throughout. The only other components we need to check are $S_{ij}$ and $S_{i0}$. We now compute 
\bea
\partial_t S_{ij} = \partial_t X_{ij} - \partial_t (\partial_i C_j) = \partial_i X_{0j} - \partial_t (\partial_i C_j) = \partial_i \partial_t C_j - \partial_t \partial_i C_j = 0
\eea
where above we use the EOM that we evolve, $\partial_t X_{ij} - \partial_i X_{0j}=0$, and the other evolved EOM, $\partial_t C_i = X_{0i}$. A similar analysis holds for 
\bea
\partial_t S_{i0} = \partial_t X_{i0} - \partial_t (\partial_i C_0) = \partial_i X_{00}-\partial_t (\partial_i C_0) = \partial_i \partial_t C_0 - \partial_t \partial_i C_0 = 0
\eea
where above we use the EOM that we evolve, $\partial_t X_{i0} - \partial_i X_{00}=0$, and the other evolved EOM, $\partial_t C_0 = X_{00}$. Therefore, if we use initial data for which all elements of $S_{\mu\nu}$ are zero, then they will remain zero throughout the evolution. This means that the energy-momentum tensor components $T^{00}$, $T^{0i}$, and $T^{ij}$ we reconstruct are indeed the ones from BDNK and the solution of our system of equations \eqref{EOM_vec} is a solution of the original BDNK equations for the chosen initial data. 

It is worth mentioning that the original dynamical variables are $T$ and $u^i$. However, since the transformation $C^\mu=T u^\mu$ is invertible (provided $C^\mu$ is timelike) with inverse $T=\sqrt{-C^\mu C_\mu}$ and $u^i=C^i/\sqrt{-C^\mu C_\mu}$, the condition for the equivalence of the solutions is to correctly define $C^\mu=T u^\mu$ at $t=0$. With this initial condition, the two systems --- the original one for $T$ and $u^i$ and the new one for $C^\mu$ --- are equivalent by a simple change of variables.

\subsection{Cleaning fields} \label{section:CurlCleaning}

In this section, we implement the GLM method to handle the dynamical evolution of the constraints present in the theory. The idea originated with Dedner et al.~\cite{DEDNER2002645} and consists of adding artificial auxiliary dynamical variables and modifying some of the dynamical equations of motion. These new fields are then forced to obey hyperbolic damping evolution equations that automatically lead to the preservation of the system's constraints during the evolution. Although the original idea was used to address numerical errors that grow exponentially in magnetohydrodynamics due to the divergence constraint $\nabla\cdot \mathbf{B}=0$, these techniques have been expanded to more general constraint equations, especially in general relativity. In particular, we follow the recipe described in Ref.~\cite{DUMBSER2020109088}, which we explain in what follows.

Once we introduce the new variables via \eqref{EOM_Cm}, we are led to the identity $[\partial_\mu,\partial_\nu]C_\alpha=\partial_\mu X_{\nu\alpha}-\partial_\nu X_{\mu\alpha}=0$, which may be split into the equations
\bml
\label{Cleaning1}
\begin{align}
\partial_0 X_{i\alpha}-\partial_i X_{0\alpha}&=0,\label{Cleaning_1a}\\
\epsilon_{ijk}\partial_j X_{k\alpha}&=0.\label{Cleaning_1b}
\end{align}
\eml
In particular, the 12 auxiliary equations \eqref{Cleaning_1a} are dynamical equations for $X_{i\alpha}$ and are, therefore, evolved in time correctly. However, the remaining 12 equations \eqref{Cleaning_1b} are not dynamical but rather constraints for the new variables. Due to numerical errors, the variables $X_{i\alpha}$ are approximately equal to $\partial_i C_\alpha$ but not exactly the same. Therefore, the zero curl constraint \eqref{Cleaning_1b} is, in fact, not preserved during the simulation. This is known to cause severe numerical behavior such as exponential growth in numerical errors away from 1+1 dimensions. To address this, we introduce auxiliary fields $\varphi_{i\alpha}$ for each constraint to ``curl clean'' the system. For each three-vector $\bm{\varphi}_\alpha=(\varphi_{1\alpha},\varphi_{2\alpha},\varphi_{3\alpha})$, we also need to introduce ``divergence cleaning'' fields $\phi_\alpha$ to control the divergence of $\bm{\varphi}_\alpha$. The details may be found in the references above; we quote the resulting equations here:
\bml
\begin{align}
\partial_0 X_{i\alpha}-\partial_i X_{0\alpha}+\textcolor{blue}{\epsilon_{ijk}\partial_j \varphi_{k\alpha}}&=0,\label{Cleaning_2a}\\
\textcolor{blue}{\partial_0\varphi_{i\alpha}}-a_\alpha^2\epsilon_{ijk}\partial_j X_{k\alpha}+\textcolor{red}{\partial_i\phi_\alpha}&=\textcolor{blue}{-\frac{\varphi_{i\alpha}}{\tau_{1\alpha}}},\label{Cleaning_2b}\\
\textcolor{red}{\partial_0\phi_\alpha}+\textcolor{blue}{b_\alpha^2 \partial_i\varphi_{i\alpha}}&=\textcolor{red}{-\frac{\phi_\alpha}{\tau_{2\alpha}}},\label{Cleaning2_c}
\end{align}
\eml
where $\epsilon_{ijk}$ is the skew-symmetric Levi-Civita pseudo-tensor, the blue terms are the corrections from the curl cleaning fields, while the terms in red are the corrections from the divergence cleaning fields. One may show that the above equations combine to form damped wave-like equations that relax to zero cleaning fields, leading the solution asymptotically to \eqref{Cleaning1} --- see \cite{DEDNER2002645,DUMBSER2020109088} for details. The constants $\tau_{1\alpha}$ and $\tau_{2\alpha}$ are positive relaxation times for the cleaning fields, while $a_\alpha$ and $b_\alpha$ are the constant propagation speeds of the cleaning fields. Because these fields are artificial, they may actually have superluminal velocities to outpace the spread of numerical errors in the form of constraint violations. To balance damping and propagation, it has been proposed that the ratio $\tau^{-1}\sim c_h/\Delta x$ \cite{DEDNER2002645}, where $\Delta x$ denotes the grid size. In applications such as relativistic magnetohydrodynamics \cite{etienne2012illinoisgrmhd}, the optimum values for the cleaning field velocities were found to be in the range between 1 and 2. On the other hand, in applications involving black holes with extreme curvatures, the optimum results were obtained for velocities $\gg 1$ \cite{alic2012z4}. These examples show the importance of adapting the optimum values of these constants in accordance with the application.

With the introduction of the fields $\phi_\alpha, \varphi_{i\alpha}$, we now have the set $\{ T^{0\nu}, C_\nu, X_{i\alpha},\varphi_{i\alpha}, \phi_\alpha\}$ of 36 dynamical variables in three spatial dimensions. The equations of motion can again be expressed in flux-conservative form, analogous to \eqref{EOM_CL}, and read
\be
\partial_0\begin{pmatrix}
T^{0\nu}\\
C_\nu\\
X_{i\alpha}\\
\varphi_{i\alpha}\\
\phi_\alpha
\end{pmatrix}+\partial_j
\begin{pmatrix}
T^{j\nu}\\
0_{4\times 1}\\
-\delta^j_i X_{0\alpha}+\epsilon_{ijk}\varphi_{k\alpha}\\
-a_\alpha^2 \epsilon_{ijk}X_{k\alpha}+\delta_i^j\phi_\alpha\\
b_\alpha^2\varphi_{j\alpha}
\end{pmatrix}
=\begin{pmatrix}
-\Gamma_{\mu\lambda}^\mu T^{\lambda\nu} - \Gamma_{\mu \lambda}^\nu T^{\mu\lambda}\\
X_{0\nu}\\
0_{12\times 1}\\
-\frac{\varphi_{i\alpha}}{\tau_{1\alpha}}\\
-\frac{\phi_{\alpha}}{\tau_{2\alpha}}
\end{pmatrix}.
\ee
The initial conditions must be set in accordance with the equivalence of the equations discussed previously, where \eqref{EOM_Cm} must hold, and, additionally, the cleaning fields must all be set initially to zero.

We remind the reader, however, that in this work we focus on 1+1-dimensional dynamical simulations in 4-dimensional Minkowski spacetime with Cartesian coordinates, so all dynamical fields depend only on $t$ and $x$. Thus, we emphasize that curl cleaning is not needed in this case, and one can simulate the original equations of motion in \eqref{EOM_vec}, where we now have only six evolved variables defined by $\mathbf{q} = (T^{00},T^{0x},C_0,C_x,X_{x0},X_{xx})^T$. This is the approach adopted in this work.

\subsection{Parameters in conformal BDNK hydrodynamics} \label{section:conformalBDNK}

Though the flux-conservative formulation developed in Section \ref{section:flux-conservative-formulation} is general and can, thus, be applied in nonconformal fluids, for the sake of simplicity, in this work we focus on conformal BDNK theory with no baryon current \cite{Bemfica:2017wps}. Conformal symmetry implies that the trace of the energy-momentum tensor is zero, $T^{\mu}_\mu = 0$, and as such, $P = \varepsilon/3$ and $c_s^2=1/3$. In the conformal regime, the energy density is given by $\ep = \alpha T^4$, where $\alpha$ is a numerical coefficient ($\alpha = 10$ in our simulation, as in \cite{Pandya_2021}), the entropy is $s = 4P/T$, $\eta/s$ is a constant, and the bulk viscosity $\zeta$ vanishes. The BDNK transport parameters $\tau_{\ep}$ and $\tau_\mathcal{Q}$ must scale inversely with the temperature, and we define them as follows \cite{Bemfica:2017wps}
\begin{equation}\label{tauep-tauQ-conformal}
    \tau_{\ep} = a_1 \frac{1}{T} \frac{\eta}{s}, \quad \tau_\mathcal{Q} =  a_2 \frac{1}{T} \frac{\eta}{s},
\end{equation}
where $a_1$ and $a_2$ are constants. We can now write the out-of-equilibrium correction to the pressure in the conformal case as one-third the energy density correction, $\Pi = \mathcal{A}/3$, so that altogether the conformal energy-momentum tensor for BDNK has the form \cite{Bemfica:2017wps}
\bea \label{Tmn_conformal}
T^{\mu\nu} = (\ep + \cA) \left ( u^\mu u^\nu + \frac{\Delta^{\mu\nu}}{3} \right )
- 2 \eta \sigma^{\mu\nu} + u^\mu \cQ^\nu + u^\nu \cQ^\mu.
\eea
As seen previously, the stress tensor can be expressed in terms of the susceptibility tensor, $H^{\mu\nu\alpha\beta}$, and the explicit expression of it in the conformal case is given by \eqref{defineBDNKnew} with $c_s^2 = 1/3$ and $\zeta=0$.

\subsection{Hydrodynamic frames used in this work} \label{section:differences}

In this work we investigate how changes in hydrodynamic frames affect the numerical solutions by choosing a set of three parameters $a_1$, $a_2$, and $c_+$ within the causal and stable regime of validity of conformal BDNK, represented by the following relations
\bea \label{BDNK_a1a2cplus}
a_1 \geq 4, \quad a_2 \geq \frac{3 a_1}{a_1-1}, \quad c_+^2 = \frac{a_1(2+a_2) + 2 \sqrt{a_1(a_1+a_1 a_2 + a_2^2)}}{3 a_1 a_2},
\eea
where $c_+$ sets the maximum characteristic velocity. Relations \eqref{BDNK_a1a2cplus} follow directly from \eqref{Causality_conformal} and \eqref{Characteristics_conformal}. The specific choices of hydrodynamic frames used in this work are defined in Table \ref{tab:hydro_frames}. The first hydro frame, $F_1$, where the maximum propagation speed is set by the characteristic $c_+ = 1$, is sharply causal, whereas the remaining two frames, $F_2$ and $F_3$, are strictly causal with $0<c_+ < 1$ \cite{Freistuhler2021}. The maximum characteristic speed, as well as all other characteristic speeds, can be set by choices of the frame parameters $a_1$ and $a_2$. 

We solve the equations of motion in the different hydrodynamic frames and compare the corresponding solutions. Since we use initial conditions where $\mathcal{A}=0$ and $\mathcal{Q}^{\mu}=0$ at $t=0$ (see Sec.\ \ref{section:IC}), the initial energy-momentum tensor $T^{\mu\nu}$ is uniquely determined by the initial temperature $T$ and flow velocity $u^{\mu}$, independently of the BDNK frame parameters $a_1$ and $a_2$. Therefore, for this class of initial data, we expect identical evolution across different causal frames when the solution remains within the regime of validity of first-order BDNK theory.

When significant disagreements between the different frames are found, we interpret this as evidence that the numerical solution is evolving outside the regime of validity of \textit{first-order} BDNK theory~\cite{Bea:2023rru}. (We note that for generic initial data with non-vanishing $\mathcal{A}$ or $\mathcal{Q}^{\mu}$, different frames would naturally lead to different dynamical evolutions even within the theory's validity regime, as the initial $T^{\mu\nu}$ itself differs from frame to frame.) On the other hand, the case where the solutions depend very weakly on the choice of causal and stable frames defines the \textit{hydrodynamic-frame robust} regime, which we focus on in this work.

\begin{table}[]
\centering
\begin{ruledtabular}
\begin{tabular}{lccc}  

 \hline
 \bf Hydro Frame & $a_1$ & $a_2$ & $c_+$ \\ [1ex] 
 \hline\hline
 $F_1$ & 25/4 & 25/7 & 1.0 \\
 \hline
 $F_2$ & 25/2 & 25/3 & 0.85 \\
 \hline
 $F_3$ & 25 & 25 & 0.74 \\ 
 \hline
\end{tabular}
\end{ruledtabular}
    \caption{Hydrodynamic frames used in the simulations of a conformal BNDK fluid.}
    \label{tab:hydro_frames}
\end{table}

\section{Numerical Implementation in 1+1 Cartesian Dynamics} \label{section:numericalimplementation}

In the present work we focus on 1+1 simulations of conformal BDNK hydrodynamics in 4-dimensional Minkowski spacetime with Cartesian coordinates. The dynamical fields in $\beta_\mu$ are taken to be $T = T(t,x)$ and $u^\mu(t,x) = (\gamma(t,x),u^x(t,x),0,0)$. As such, we can write $C_\mu = (C_0, C_x, 0, 0)$ and $u^0 = \gamma = \sqrt{1 + u_x^2}$, and only $X_{00}$, $X_{0x}$, $X_{x0}$, and $X_{xx}$ are nonzero. The general equations of motion can be succinctly written as
\begin{equation}
	\partial_t \mathbf q + \partial_x \mathbf F^x = \mathbf S,
\end{equation}
or, explicitly:
\begin{equation}
	\partial_t \m{
		T^{00} \\ T^{0x} \\ C_0 \\ C_x \\ X_{xx} \\ X_{x0}
	} +
    \partial_x \m {
        T^{x0} \\ T^{xx} \\ 0  \\ 0 \\ - X_{0x} \\ - X_{00}
    } =
    \m {
        0 \\ 0 \\ X_{00} \\ X_{0x} \\ 0 \\ 0
    }.
    \label{EOM1+1}
\end{equation}
Note that only the components $X_{xx}$ and $X_{x0}$ of $X_{\mu\nu}$ are evolved variables, whereas the other two components $X_{0x}$ and $X_{00}$ must be calculated from the evolved $T^{00}$ and $T^{0x}$ components of the energy-momentum tensor, by inverting a simple $2 \times 2$ matrix obtained from \eqref{defineTnew}, as calculated in \eqref{X_Inverse}. 

In order to solve the equations of motion, we use a standard Total Variation Diminishing Runge-Kutta of second order (TVDRK2) \cite{Shu:1988iw}: we integrate in time from step $l$ to $l+1$ with intermediate terms indexed by superscript $(1)$, seen here
\eq{
\mathbf q^{(1)} = \mathbf q_l - h_t \mathbf Z(\mathbf q_l)
}
\eq{
\mathbf q_{l+1} = \mathbf q^{(2)} = \frac{1}{2} \mathbf q_l + \frac{1}{2} \mathbf q^{(1)} - \frac{1}{2} h_t \mathbf Z( \mathbf q^{(1)} ).
}
The spatial residual is defined as $\mathbf Z(\mathbf q) = \partial_x \mathbf F^{x}(\mathbf q) - \mathbf S(\mathbf q)$ and the RK coefficients are $\alpha_{ij} = \{\alpha_{10} = 1, \alpha_{20} =1/2, \alpha_{21} = 1/2 \}, \beta_{ij} = \{\beta_{10}= 1, \beta_{20} = 0, \beta_{21} = 1 \} $. Therefore, we find the bound on the CFL coefficient \cite{1967_courant_friedrichs_lewy} to be $0 < c_{\text{CFL}} \leq 0.5$, relating the temporal and spatial resolution in our simulations according to $h_t \leq c_{\text{CFL}} h_x$. We employ the well-known Kurganov-Tadmor algorithm \cite{KurganovTadmor} for the spatial evolution, as reviewed in the following section.

\subsection{Kurganov-Tadmor algorithm} \label{section:KT}

In this section, we give some details of the algorithms used in this work. A finite-volume spatial evolution algorithm that can handle shocks very well is the second-order central scheme known as Kurganov-Tadmor (KT), which is commonly used in hydrodynamic simulations of the quark-gluon plasma, see \cite{Jeon:2015dfa,Romatschke:2017ejr}. An extension of the Nessyahu-Tadmor scheme, this algorithm retains a lack of dependence on a problem's eigenstructure and boasts a much smaller numerical viscosity that is independent of temporal resolution, allowing for a simple semi-discrete formulation \cite{KurganovTadmor,BazowKT}.
The semi-discrete update equation is given by
\eq{
 \frac{d}{d\tau} \mathbf q_{i j k} = - \sum_{n=1}^3 \frac{ \mathbf H_{i+1/2,j,k}^{x_n} - \mathbf H_{i-1/2,j,k}^{x_n}}{\Delta x_n}+ \mathbf S[ \mathbf q_{i j k}], \quad  (x_1 = x, x_2 = y, x_3 = z )
}
 where the right-hand side is defined as the residual, $\mathbf Z[ \mathbf q]$, and includes the numerical fluxes $\mathbf H$ and source $\mathbf S$. The index $n$ sums over all spatial dimensions $x_n$. The indices $ijk$ are the integer labels for spatial coordinates of a grid point, and $\Delta x_n$ are the numerical resolution in spatial coordinates. Numerical fluxes appear in the update equation as a way of describing the time evolution of the conserved variable vector $\mathbf q$. They are
\bea \label{KTflux}
\mathbf H_{i\pm 1/2,j,k}^{x_n} \equiv \frac{  \mathbf F^{x_n}[\mathbf q^+_{i\pm1/2,j,k}] +  \mathbf F^{x_n}[\mathbf q^-_{i \pm1/2,j,k}]}{2} - a^{x_n}_{i\pm 1/2,j,k} \frac{ \mathbf q^+_{i\pm1/2,j,k} - \mathbf q^-_{i \pm1/2,j,k}}{2}\,.
\eea
KT stands out due to its ability to be more precise with the information on the propagation speed of the system by making use of the local propagation speed at each cell interface \cite{KurganovTadmor}, given by
\eq{ a^{x_n}_{i\pm 1/2,j,k} \equiv \text{max} \left \{ \rho \left ( \frac{\partial \mathbf F^{x_n}}{\partial \mathbf q} [\mathbf q^+_{i\pm 1/2,j,k}] \right ), \rho \left( \frac{\partial \mathbf F^{x_n}}{\partial \mathbf q} [\mathbf q^-_{i \pm1/2,j,k}]
\right )
\right \},
}
where $\rho(A) \equiv \text{max}_i(| \lambda_i(A)| )$ is the spectral radius and $\lambda_i(A)$ are the eigenvalues of the Jacobian matrix $A \equiv \partial \mathbf F / \partial \mathbf q$. Here, $+$ and $-$ superscripts stand for reconstructed values of $\mathbf q$ at the left and right sides of the corresponding numerical cells $i+1/2$ and $i-1/2$, respectively. Looking at the $x$ spatial dimension, we have intermediate values
\begin{align}
& \mathbf q^{+}_{i + 1/2,j,k} \equiv \mathbf q_{i + 1,j,k} - \frac{\Delta x}{2} ( \mathbf q_{x})_{i + 1,j,k}, \quad \mathbf q^{-}_{i + 1/2,j,k} \equiv \mathbf q_{i,j,k} + \frac{\Delta x}{2} ( \mathbf q_{x})_{i,j,k} \\
& \mathbf q^{+}_{i - 1/2,j,k} \equiv \mathbf q_{i,j,k} - \frac{\Delta x}{2} (\mathbf q_{x})_{i,j,k}, \quad
\mathbf q^{-}_{i - 1/2,j,k} \equiv \mathbf q_{i-1,j,k} + \frac{\Delta x}{2} (\mathbf q_{x})_{i-1,j,k}\,. 
\end{align}

The approximate spatial derivatives $(\mathbf q_{x})_{i,j,k}$ are constructed to satisfy the ``total variation diminishing'' (TVD) property, which prevents the introduction of new local extrema in the numerical solution. In practice, TVD means that the scheme avoids spurious oscillations near sharp features such as shocks or contact discontinuities while remaining high-order accurate in smooth regions. This balance is achieved by introducing flux limiters that enforce monotonicity in the reconstruction step. The specific flux limiters are implemented through the method-dependent flux parameter $\phi(r)$, which depends on the ratio
\bea
r = \frac{\mathbf q_i-\mathbf q_{i-1}}{\mathbf q_{i+1} - \mathbf q_i + \epsilon},
\eea
where $\epsilon$ is a small tolerance included to avoid division by zero. The one-dimensional spatial gradient then takes the form
\bea
(\mathbf q_x)_i = \phi(r_i) \frac{\mathbf q_{i+1} - \mathbf q_i}{h_x}.
\eea

\subsubsection{Generalized minmod limiter}
The minmod limiter \cite{KurganovTadmor} for numerical derivatives guarantees the TVD non-oscillatory property by satisfying a local scalar maximum principle, 
\bea
\phi(r_i) = \minmod \left (\theta r_i, \frac{1+r_i}{2}, \theta \right )
.
\eea
 This gives a spatial derivative,
\begin{equation}
    (\mathbf q_x)_{i,j,k} \equiv \minmod\left(\theta \frac{\mathbf q_{i,j,k} - \mathbf q_{i-1,j,k}}{\Delta x},\frac{\mathbf q_{i+1,j,k} - \mathbf q_{i-1,j,k}}{2 \Delta x}, \theta \frac{\mathbf q_{i+1,j,k} - \mathbf q_{i,j,k}}{\Delta x}\right)\,.
\end{equation}
The parameter $\theta \in [1,2]$ controls the amount of numerical dissipation: $\theta =1$ corresponds to the most dissipative limiter, while $\theta = 2$ yields the least dissipative limiter. This multivariable minmod limiter can be expressed in a more illuminating way by breaking it into parts,
\begin{equation}
    \minmod(x,y,z) \equiv \minmod(x,\minmod(y,z)),
\end{equation}
where it performs this operation $\minmod(x,y) \equiv [\sgn(x) + \sgn(y)] \min(|x|,|y|)$ and $\sgn(x) = |x|/x $. We use a dissipation parameter of $\theta = 1$ in all the simulations reported in this work.

Although the above provides the theoretical framework, the actual implementation follows directly by applying these limiter-based reconstructions at each cell interface during the numerical update. The important point is that KT, paired with the generalized minmod limiter, yields a scheme that is both robust near shocks and accurate in smooth regions \cite{KurganovTadmor}.

\section{Numerical results} \label{section:Results}

We now present numerical results for our new formulation of BDNK hydrodynamics. We employ two different sets of initial conditions and investigate convergence and frame robustness for different values of the shear viscosity to entropy density ratio.

To assess numerical convergence, we monitor the scaling of numerical errors with the grid spacing $h$. 
For a second-order method such as Kurganov-Tadmor, these errors are expected to scale as $|\delta q| \sim h^2$ for asymptotically small $h$ \cite{KurganovTadmor}. 
The scaling power is estimated, as a function of time, by the convergence function $Q(t)$, defined in Eq.\ \eqref{convergence_test}, 
which is therefore expected to approach the value of $2$ in the convergent regime. Details can be found in Appendix~\ref{appendix:convergence}.

\subsection{Setting up the initial conditions} \label{section:IC}

In this paper, we define quantities that allow us to estimate the Knudsen number \textrm{Kn} in our BDNK simulations. We opt here for two definitions, one in terms of velocity and another in terms of temperature,
\be \label{define_Kn}
\text{Kn}_u = \frac{\eta}{s} \frac{\partial_x u_x}{T},
\quad
\text{Kn}_T = \frac{\eta}{s} \frac{\partial_x T}{T^2}.
\ee
Expressed in terms of variables from our flux-conservative formulation \eqref{EOM_vec}, they become
\be
\text{Kn}_u = \gamma^2 \left ( \frac{\eta}{sT^2} X_{xx} - v \frac{\eta}{s T^2} X_{x0}
\right ),
\quad
\text{Kn}_T = \gamma \left (\frac{\eta}{sT^2} X_{x0} - v  \frac{\eta}{sT^2} X_{xx} \right ).
\ee
Here, the microscopic scale is taken to be $\eta/(sT)$, and the scale of variation of the conserved quantities is estimated using $\partial_x u_x$ and $\partial_x T/T$.
Since the energy-momentum tensor of BDNK is defined by a first-order truncation of the primary hydrodynamic variables, in its regime of validity, deviations from equilibrium must remain small. This can be monitored by the Knudsen numbers defined above, which should start and remain small in simulations that are within the domain of validity of the \emph{first-order} BDNK theory.

The set of initial conditions that will be used throughout this work describes a system that initially has vanishing $\mathcal{A}$ and $\mathcal{Q}^\mu$, but which can still have nonzero shear stress $\pi^{\mu\nu}\neq 0$. Therefore, for this particular class of initial data, the only dissipative term that can contribute to the initial full $T^{\mu\nu}$ is the contribution coming from $\sigma^{\mu\nu}$. Thus, the BDNK evolution starts as close as possible to a Navier-Stokes state, which can be compatible with the regime of validity of the \emph{first-order} BDNK formulation.

Our initial data is a convenient choice to study the possibility of shock formation in BDNK. In fact, in this case, one can choose initial data for $T$ and $u^\mu$ that leads to shocks for the corresponding relativistic Euler equations, and define their time derivatives needed for BDNK evolution by the condition that $\cA$ and $\cQ^\mu$ are zero in the initial state. This procedure was worked out in detail in \cite{Bemfica:2017wps} and here we only quote the result (for 3+1 flows, for generality), reproduced below
\bea
\partial_0 \ep |_{t=0} = \frac{4 \ep \gamma}{3 + 2 (\overline u)^2} \left ( \frac{u^l u^m \partial_l u_m}{1 + (\overline u)^2} - \frac{\partial_l u^l - u^l \partial_l \ep}{2\ep}
\right )
,
\eea
\bea
\gamma\, \partial_0 u_j |_{t=0} = \left ( \gamma^2 \partial_l u^l - u^l u^m \partial_l u_m - \frac{u^l \partial_l \ep}{4 \ep}
\right ) \frac{u_j}{3+ 2 (\overline u)^2} - u^l \partial_l u_j - \frac{\partial_j \ep}{4 \ep},
\label{eq:d0uj}
\eea
where $(\overline u)^2 = u_i u^i$. 
Now, we are able to consider initial data for $\varepsilon, u_x$ in our 1+1 simulations that would lead to  shocks in the Euler limit, and determine in a consistent manner $\partial_0\varepsilon, \partial_0 u_x$ at the initial state using the equations above.  

For the sake of completeness, we list below the components of the energy-momentum tensor needed to initialize the 1+1 evolution:
\bea \label{T00_eulerlike}
T^{00} =  \frac{\ep}{3} \left ( 4 \gamma^2 - 1 \right )- 2 \eta \,\sigma^{00},
\eea
and
\bea \label{T0x_eulerlike}
T^{0x} = \frac{4}{3} \ep \gamma u^x - 2 \eta\, \sigma^{0x}.
\eea 
When the initial velocity is not homogeneous, the shear tensor has nonzero components. In this case, it is sufficient to know that 
\bea \label{shear00_eulerlike}
\sigma^{00} = -\frac{2}{3}(\gamma^2 -1 ) \partial_t u_0 + u_0 u_x \partial_x u_0 - \frac{1}{3}(\gamma^2 -1) \partial_x u_x ,
\eea 
\bea \label{shear0x_eulerlike}
\sigma^{0x} = \frac{1}{6} u_0 u_x \partial_t u_0 + \frac{1}{2}\partial_t (\gamma^2 -1) - \frac{1}{6} \partial_x u_x(u_0 u_x) - \frac{\gamma^2}{2} \partial_x u_0,
\eea
as the other nonzero components can be obtained by imposing the tracelessness and orthogonality properties of the shear tensor.
The time derivative of the velocity is then given by
Eq.~\eqref{eq:d0uj} and by the normalization condition $u^\mu u_\mu = -1$, which yields $\partial_t u_0 = -u^x \partial_t u_x/u^0$. 
Now, we have the information needed  to initialize the state $\mathbf q = (T^{00}, T^{0x}, C_0, C_x, X_{xx}, X_{x0})^T$ in a simulation.

\begin{figure}[t]
    \centering
    \includegraphics[width=1\textwidth]{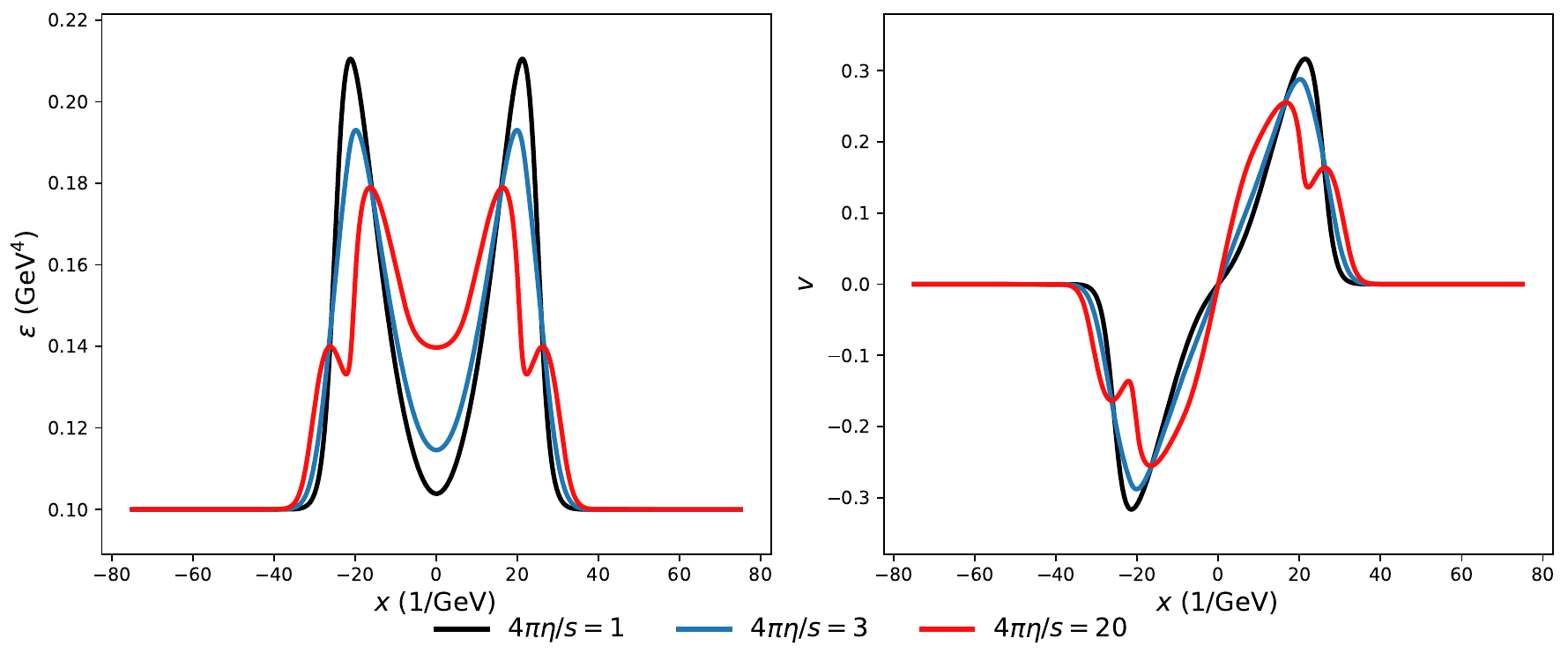}
    \caption{A snapshot of the hydrodynamic evolution at $t=30$ GeV$^{-1}$ for Gaussian initial conditions and $4\pi \eta/s = 1,3,20$. Here we employ the $F_2$ hydrodynamic frame, with $c_+ = 0.85,$ $ a_1 = 25/2, $ $ a_2 = 25/3 $. The simulations pictured were performed on a grid of 48,001 cells and a minimum energy baseline of $\ep_0 = 0.1$ GeV$^4$.}
    \label{fig1:gaussian_ep_visc}
\end{figure}
\begin{figure}[h]
    \centering
    \includegraphics[width=1.\textwidth]{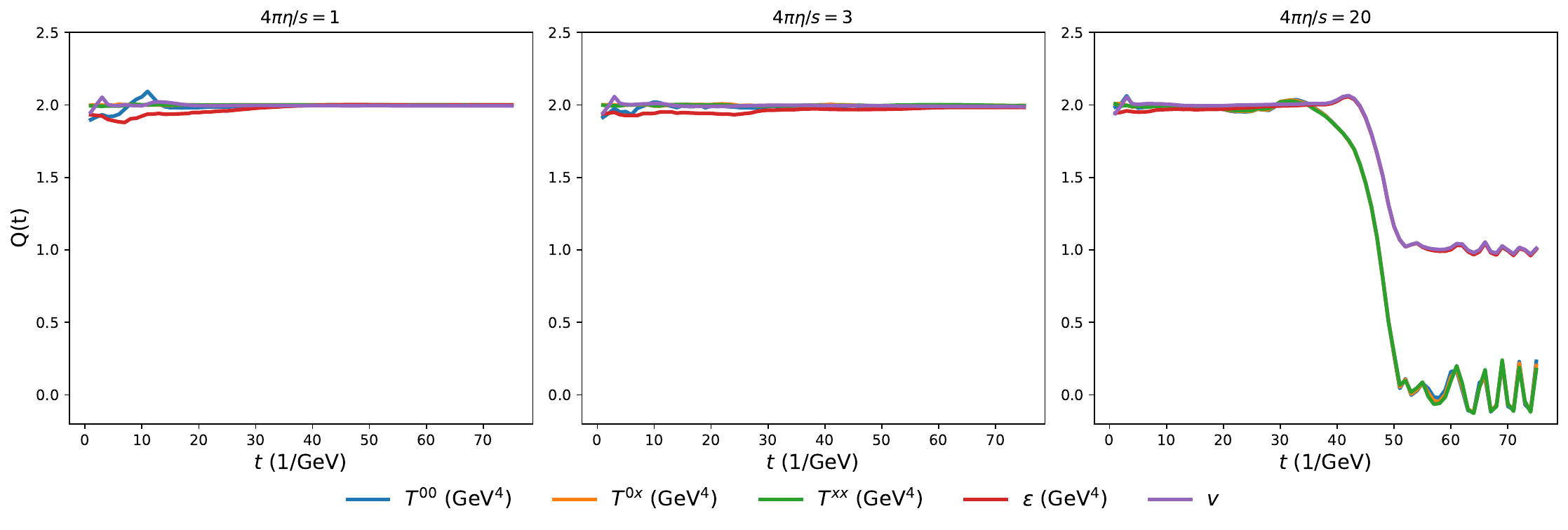}
    \caption{Convergence test using an $L_1$ norm for Gaussian initial conditions and $4\pi \eta/s = 1,3,20$. Here we employ the $F_2$ hydrodynamic frame. Simulations used in the testing had 12,001, 24,001, and 48,0001 cells.}
    \label{fig2:gaussian_convergence_delta0p1}
\end{figure}

Note that the class of initial conditions we have outlined is still quite general. Beyond the initial values for the viscous corrections in the BDNK energy-momentum tensor, we still need to provide adequate forms for the initial energy density $\ep$ (or temperature $T$) and flow $u^\mu$. 
In this first study, we have investigated 
two analytical forms for the initial energy density $\ep(t=0,x)$, while for the velocity we used $v(0,x)=0$, where $u^\mu=(\gamma, \gamma v)$ (this choice significantly simplifies the initial conditions for the new variables used in the first-order reduction, as explained below). Our first  set of initial conditions uses a Gaussian profile for the initial energy density, as defined in \cite{Pandya_2021}, while the second set uses a smooth shocktube profile for the energy density defined by a Fermi-Dirac function. We have solved these two different problems using a new open-source code\footnote{We have developed both a Python and a C++ implementation of our numerical scheme. They give the same results for the simulations shown in this paper, with the C++ version being considerably faster. The Python code can be downloaded from \url{https://github.com/clarissen/conformal-BDNK-1p1-tx} and the implementation in C++ can be found at \url{https://github.com/warduz/conformal-1p1-bdnk-cpp}. } and our results can be found in what follows.

To investigate the regime of validity of BDNK theory, the initial conditions we use must be consistent and well-behaved. As mentioned previously, \text{Kn} and the viscous corrections $\mathcal{A}$ and $\mathcal{Q}_\mu$ are good indicators of whether or not the solution is far from the theory's regime of validity. We thus consider initial data where the Knudsen number and therefore the viscous corrections are small, i.e., $\cA/(\ep+P) \ll 1$, $\sqrt{\mathcal{Q}^{\mu} \mathcal{Q}_\mu}/ (\ep +P) \ll 1$, and $\text{Kn}_{u,T} \ll 1$, typically associated with the case where $4\pi\eta/s \sim 1$. The exception to this is when we consider large values of $\eta/s$ (e.g. $\eta/s = 20/4\pi$), which is motivated by discussion found in previous works \cite{Pandya_2021,Bhambure:2024axa}.

\subsection{Gaussian initial data} \label{section:ICgaussian}

We first show results for initial data where the energy density profile is given by a Gaussian, which was analyzed before within a different numerical scheme\footnote{We have checked that our simulations for the Gaussian initial condition are in excellent agreement with the corresponding simulations with the same parameters performed in \cite{Pandya_2021}, see Appendix \ref{appendix:pandya_agreement}. We thank A.~Pandya for providing a version of their code so we could check their solutions against ours.} in Ref.\ \cite{Pandya_2021}. This profile is defined by
\bea \label{gaussianIC_ep}
    \ep(0,x) = \ep_0 + A e^{-x^2/\Delta^2} .
\eea 
In this case, one assumes $v=0$ in the initial condition, so $\sigma^{\mu\nu} = 0$, which, together with the enforced conditions $\cA|_{t=0}= \cQ_\mu|_{t=0} = 0$, makes $\partial_t \ep|_{t=0}=\partial_t v|_{t=0} = 0$. 
We note that the lack of a shear contribution to the stress tensor is contingent on this particular choice of having velocity being zero in the initial condition, and not a general result.

Although there is no shear per se in the initial data, shear-viscous contributions arise as the system evolves according to the BDNK equations. In fact, increasing $\eta/s$ can greatly affect the dynamics of the hydrodynamic fields $\ep, v$ \cite{Pandya_2021,Bhambure:2024axa}. 
In particular, for very large $\eta/s$, the system quickly develops a large viscous stress and may rapidly leave the regime of validity of BDNK, depending on the $\varepsilon(0,x)$ used. 
Therefore, results in this regime must be interpreted with care, as they are very likely outside the regime of validity of first-order hydrodynamics. 

We first take Gaussian initial conditions with parameters $A = 0.4$ GeV$^4$, $\ep_0 = 0.1$ GeV$^4$, and $\Delta = 5$ GeV$^{-1}$. 
Our numerical solution of the BDNK equations is shown in Fig.~\ref{fig1:gaussian_ep_visc}, which displays a snapshot of $\ep$ and $v$ at $t= 30$ GeV$^{-1}$, for different values of $4\pi \eta/s = 1,3, 20$. 
One can see that the profiles for the hydrodynamic variables are very similar for $4\pi \eta = 1, 3$ and significantly different for $4\pi \eta/s = 20$.
From this observation alone, one might conceive that cranking up the value of $\eta/s$ causes  issues that indicate that BDNK is unfit for extremely viscous systems. However, such a conclusion would be premature, as we discuss below.

\begin{figure}
    \centering
    \includegraphics[width=1\textwidth]{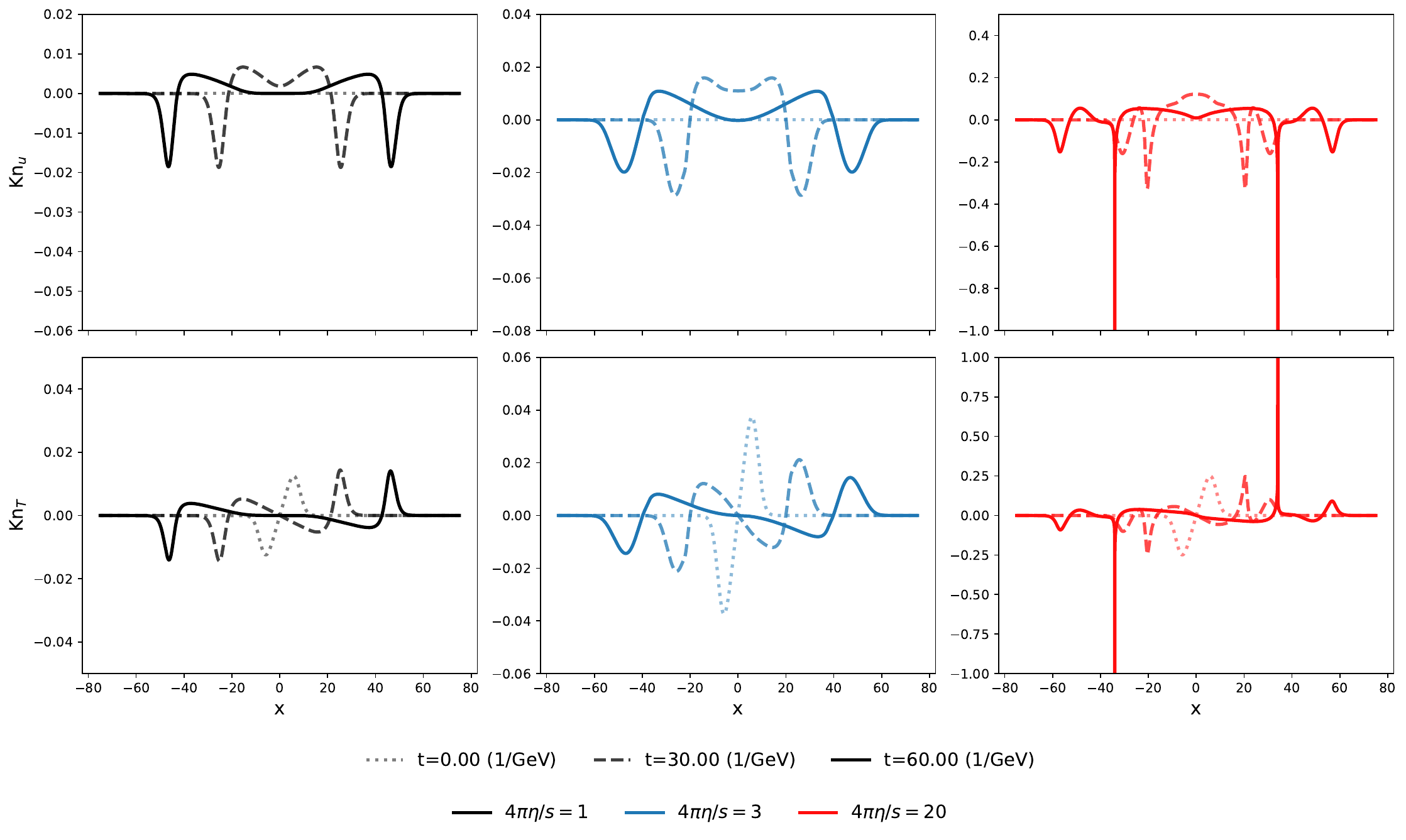}
    \caption{Three snapshots of the evolution of the  Knudsen numbers $\text{Kn}_u$ and $\text{Kn}_T$ at $t=0,30,60$ GeV$^{-1}$, 
    for Gaussian initial conditions and  $4\pi \eta/s = 1,3,20$. Here we employ the $F_2$ hydrodynamic frame. The simulations were performed on a grid of 48,001 cells, with a minimum energy density of $\varepsilon_0 = 0.1$ GeV$^4$.  }
    \label{fig3:gaussian_Kn_visc}
\end{figure}

Convergence tests for the same solutions are shown in Fig.~\ref{fig2:gaussian_convergence_delta0p1} --- see Appendix \ref{appendix:convergence} for a detailed discussion of how convergence is tested and how the convergence function, $Q(t)$, is defined. For moderate values of $\eta/s$, convergence is under control as $Q(t)$ remains nearly at the expected value of 2, especially for the hydrodynamic fields $\varepsilon$ and $v$. However, as one increases $\eta/s$ towards the much larger value of $4\pi\eta/s=20$, which was used before in \cite{Bhambure:2024axa}, $Q(t)$ significantly departs from the expected value  at late times, even for $\varepsilon$ and $v$. More importantly, $Q(t)$ falls dramatically towards zero, with fluctuations that become negative, indicating that numerical convergence is lost. 
Differences between the convergence function for $\ep$, $v$, and for the elements of $T^{\mu\nu}$ may be partially explained by the fact that the latter contains gradients of $\varepsilon$ and $v$. 
The loss of numerical convergence at late times greatly affects the Knudsen numbers \eqref{define_Kn} monitored in the simulation, shown in Fig.\ \ref{fig3:gaussian_Kn_visc}. While for moderate values of $\eta/s$ the Knudsen numbers remain small, the highest value of $4\pi\eta/s=20$ leads to very sharp peaks with $|\text{Kn}_{u,T}| \gg 1$. In fact, both $|\text{Kn}_{u,T}| > 20$, certainly well beyond acceptable bounds of a Knudsen number for a first-order effective theory. Therefore, for this type of initial data for the dynamical fields, and such very large values of $\eta/s$, the lack of convergence and the large magnitude of Knudsen numbers indicate that conclusions concerning the properties of BDNK under these extreme conditions are not reliable.

\begin{figure}
    \centering
    \includegraphics[width=1\textwidth]{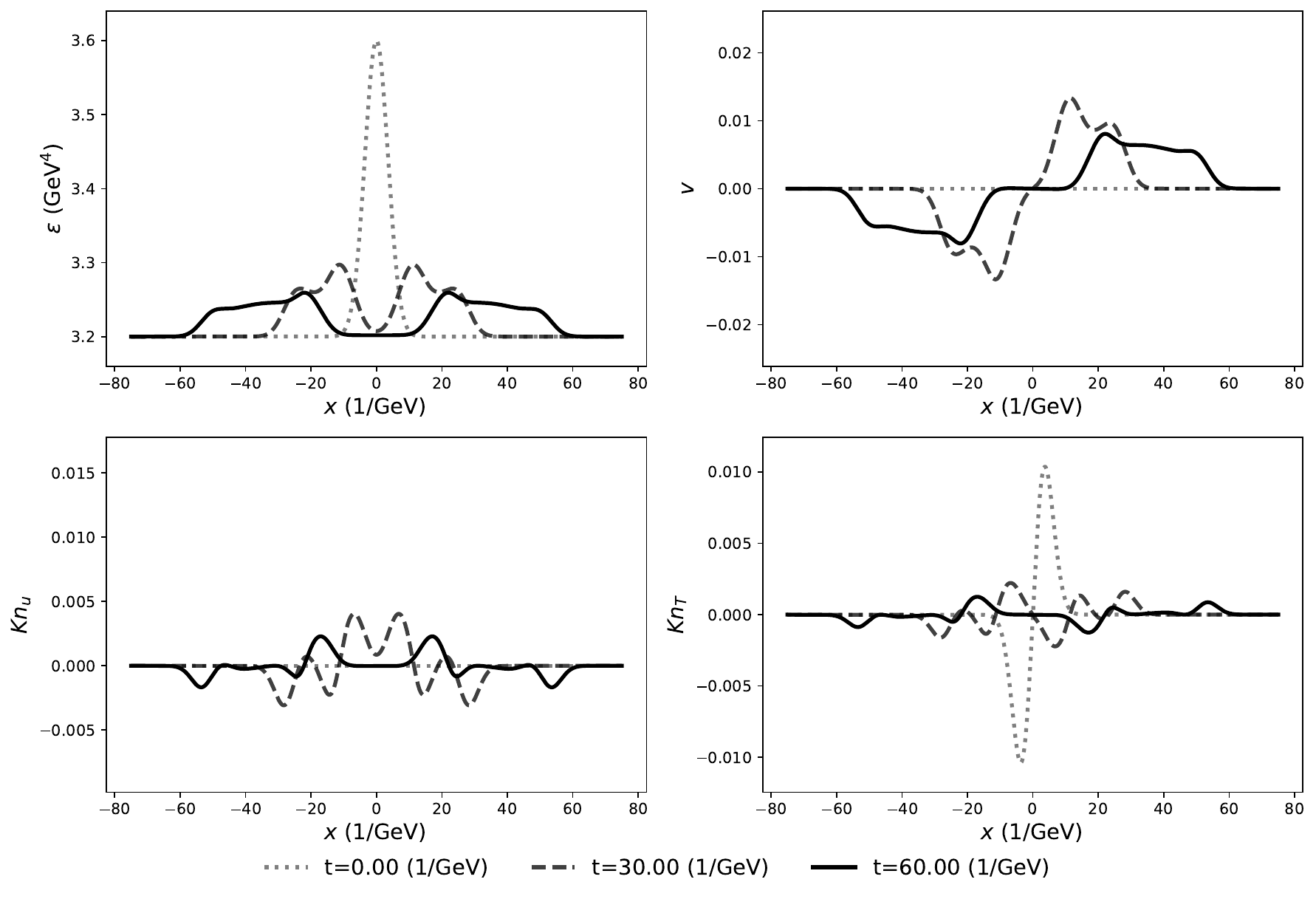}
    \caption{Three snapshots of the evolution of the  Knudsen numbers $\text{Kn}_u$ and $\text{Kn}_T$ at $t=0,30,60$ GeV$^{-1}$, 
    for Gaussian initial conditions and $4\pi \eta/s = 20$. Here we employ the $F_2$ hydrodynamic frame. This simulation used a minimum energy density of  $\varepsilon_0 = 3.2$ GeV$^4$.}
    \label{fig4:gaussian_eta=20_d=64_Kn}
\end{figure}

\begin{figure}
    \centering
    \includegraphics[width=0.5\textwidth]{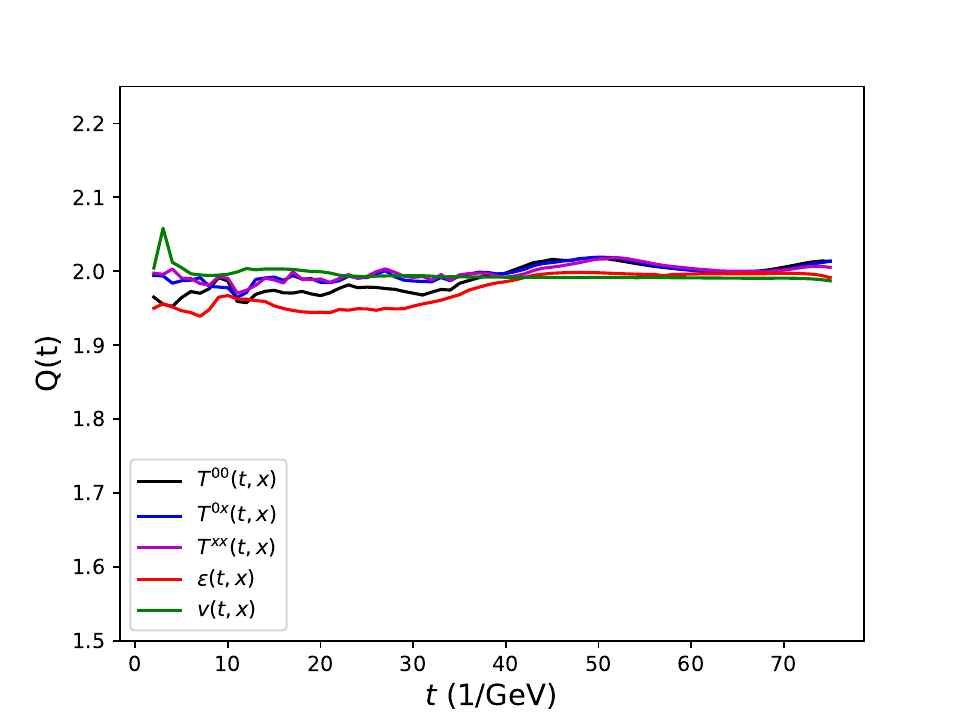}
    \caption{Convergence test using an $L_1$ norm 
    for Gaussian initial conditions with a minimum energy density of  $\varepsilon_0 = 3.2$ GeV$^4$ and $4\pi \eta/s = 20$. 
    Here we employ the $F_2$ hydrodynamic frame. The simulations used in this testing were performed on a grid of 12,001, 24,0001, and 48,001 cells. }
    \label{fig5:gaussian_eta=20_d=64_Qt}
\end{figure}

This does not mean, however, that BDNK cannot handle very large values of $\eta/s$. In fact, the large peaks in the Knudsen numbers can be simply controlled by trivially changing the initial data by raising the baseline $\ep_0$ in the energy density profile of Eq.~\eqref{gaussianIC_ep}. 
Figure~\ref{fig4:gaussian_eta=20_d=64_Kn} shows the evolution of $\ep, v$ and the Knudsen numbers from Eq.~\eqref{define_Kn} when the larger background energy density of $\ep_0 = 3.2$ GeV$^4$  is adopted. 
Even though one considers the highest $\eta/s = 20 / 4\pi$, by simply rescaling the initial data, the Knudsen numbers remain small and well behaved. As a result, convergence is maintained, with $Q(t)\simeq 2$ throughout the evolution, as shown in
Fig.~\ref{fig5:gaussian_eta=20_d=64_Qt}. This shows that BDNK can properly handle large $\eta/s$ values, as long as the initial data is such that the gradients remain small and under control, in perfect agreement with what is expected from a first-order theory with a well-defined regime of validity.

\begin{figure}
    \centering
    \includegraphics[width=1\textwidth]{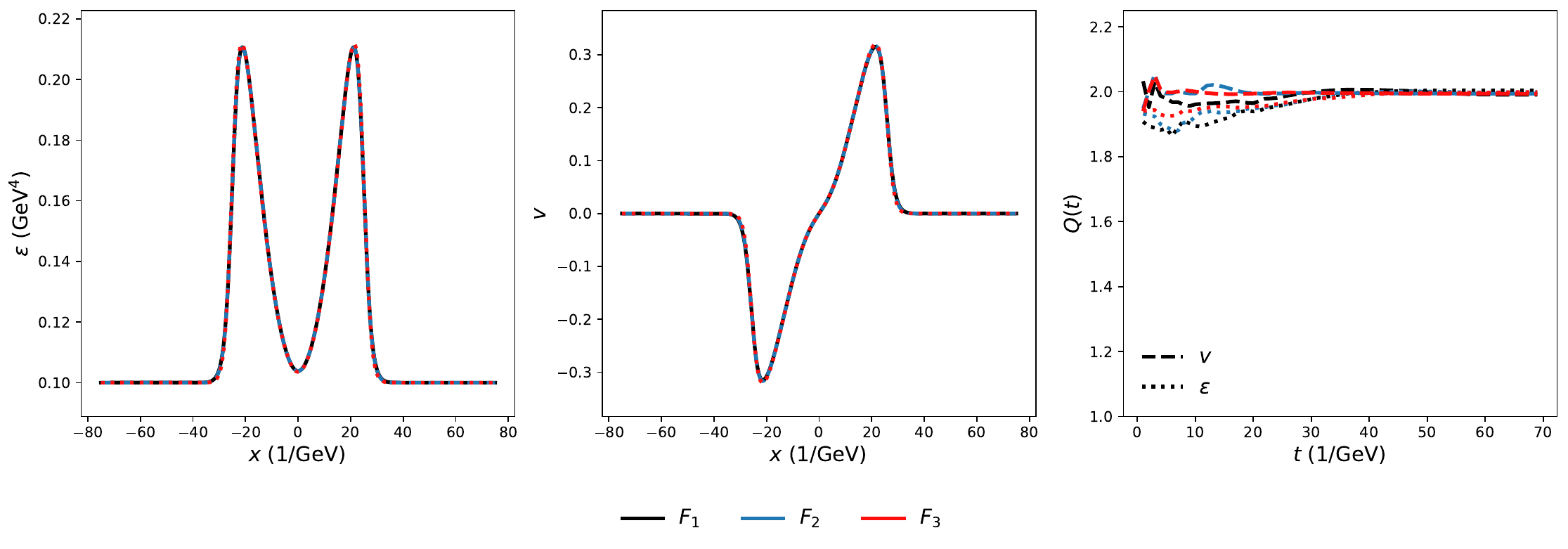}
    \caption{
    A snapshot of the hydrodynamic evolution at
    $t=30$ GeV$^{-1}$, alongside a $L_1$ convergence test  across all three hydrodynamic frames $F_1$, $F_2$ and $F_3$ in Table \ref{tab:hydro_frames},  for Gaussian initial conditions, with $4\pi \eta/s = 1$.  The simulations pictured had grids of 48,001 cells, with a minimum energy density of  $\ep_0 = 0.1$ GeV$^4$.}
    \label{fig6:gaussian_ep_v_Qt}
\end{figure}

To show that the simulations reside within the regime of validity of BDNK, we must also investigate their behavior across different hydrodynamic frames. 
We do so by comparing the three hydrodynamic frames listed in Table~\ref{tab:hydro_frames}. 
Comparisons for $\ep$ and $v$, together with the corresponding convergence tests, are included in Fig.\ \ref{fig6:gaussian_ep_v_Qt}. While there are no qualitative or visual differences to note across the evolution, there are slight variations in the convergence function across these frames. Between the two variables, $v$ more readily converges to the limit of 2, whereas $\ep$ takes a little longer across all frames. At earlier times $t<20$ GeV$^{-1}$, there is a spread in the convergence order across the frames with  the $F_2$ hydrodynamic frame, having the least spread. This frame has a maximum characteristic speed of $c_+ = 0.85$, right in the middle of the interval considered in this work, in between those for the hydrodynamic frames $F_1$ and $F_3$.

Finally, we remark that the time evolution of all components of $T^{\mu\nu}$, $C_\mu$, and $X_{\mu\nu}$ for initial conditions with a Gaussian profile for the energy density can be found in Appendix \ref{appendix:gaussian}.

\subsection{Smooth step in energy density} \label{section:ICepshock}

The second problem we studied is the case of a smooth shocktube initial condition in the energy density. Typically, one considers the Euler piece-wise initial condition that takes a left and right state distribution in the energy density
\bea \label{vLvRStep}
\ep(t=0,x) = \begin{cases}
            \ep_L, \quad x \leq 0 \\
            \ep_R, \quad x >0.
\end{cases}
\eea
However, note that these initial conditions would carry a diverging gradient $\partial_x \ep$ into the flux-conservative form we present in this paper, \eqref{EOM1+1}. To avoid such issues, we connect these two states over a continuous gradient specified by a length scale $\Delta$. We then consider the following modified distribution
\bea \label{define_epshock_IC}
\ep(t=0,x) = \ep_R + \frac{\ep_L - \ep_R}{1 + \text{e}^{x/\Delta} },
\eea
and fix the initial velocity in $x$ to be zero, $v = 0$. The left and right states of the energy density are set by $\ep_L, \ep_R = 1.3, 0.3$ GeV$^{4}$, and their separation is set by $\Delta = 1$ GeV$^{-1}$. The initial time derivatives of $\ep$ and $v$ are computed according to Sec. \ref{section:IC}, and because $v=0$, just as in the previous initial condition, we have $\partial_t \ep |_{t=0} = \partial_t v|_{t=0} = 0$. In this scenario, the initial state vector has the form $ q(t=0) = ( T^{00}_0 - 2 \eta \sigma^{00}, T^{0x}_0 - 2 \eta \sigma^{0x},  -T_0(x),  0,  0, - \partial_x T_0(x) )^T$. From this, all the flux and source elements can then be computed.

\begin{figure}[htbp]
    \centering
    \includegraphics[width=1.0\textwidth]{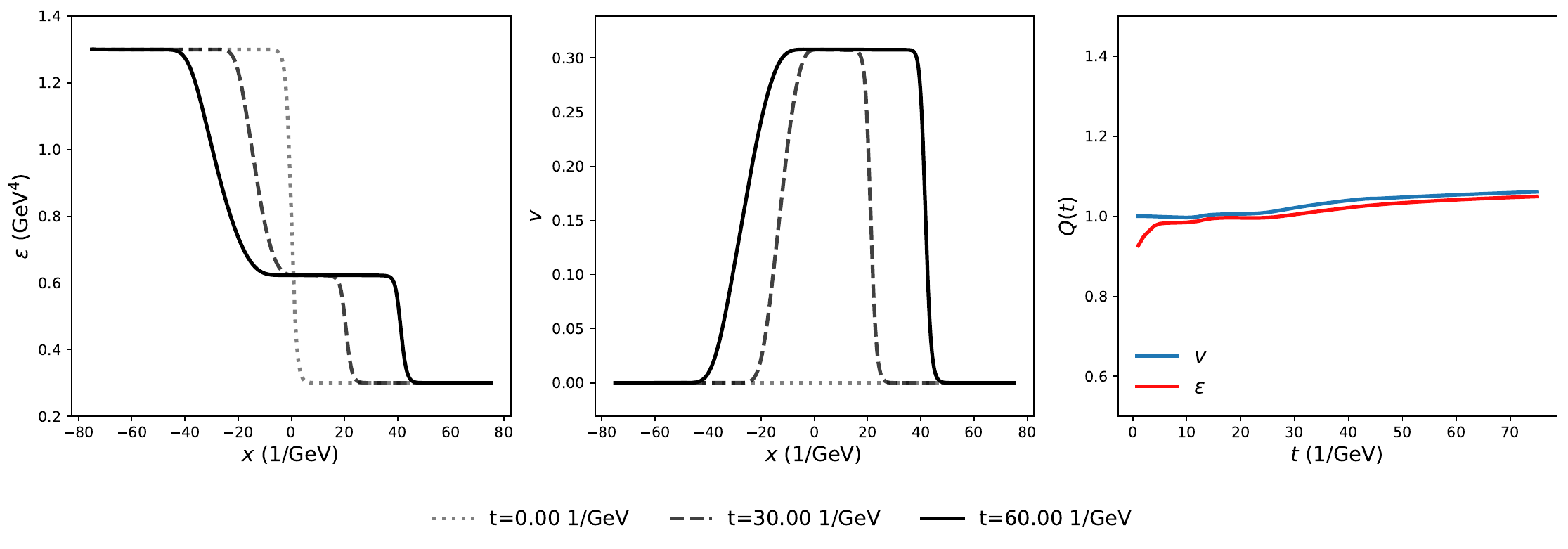}
    \caption{
    Three snapshots of the hydrodynamic evolution at
    $t=0,30,60$ GeV$^{-1}$, alongside a $L_1$ convergence test,  for stair-step initial conditions and $4\pi \eta/s = 1$. 
    Here we employ the $F_2$ hydrodynamic frame.
    The simulations pictured were performed on a grid of 48,001 cells, and additional simulations of 12,001 and 24,001 cells were used in the convergence testing.}
    \label{fig7:ep_shock_ep_v_Qt}
\end{figure}

\begin{figure}[htbp]
    \centering
    \includegraphics[width=0.5\textwidth]{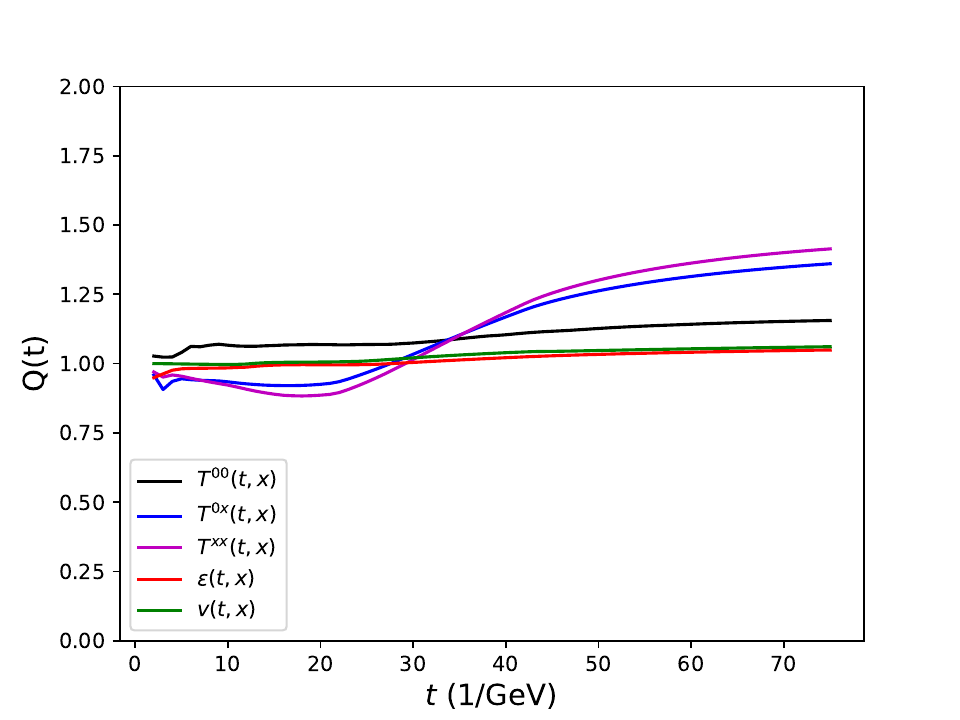}
    \caption{Convergence test using an $L_1$ norm 
    for stair-step initial conditions and $4\pi \eta/s = 1$. 
    Here we employ the $F_2$ hydrodynamic frame. The simulations were performed on a grid of 12,001, 24,001 and 48,001 cells.}
    \label{fig9:ep_shock_convergence}
\end{figure}

 Three snapshots of the evolution of $\ep, v$, at times $t= 0, 30, 60$ GeV$^{-1}$, are shown in Fig.\ \ref{fig7:ep_shock_ep_v_Qt},  alongside the convergence test $Q(t)$ for both variables, for  $\eta/s = 1/4\pi$. 
The evolution displays the left and right-moving waves, which are typical in this problem. 
While the left-moving wave rarefies with time, the right-moving wave seems to be evolving to form a shock.
The convergence order starts out and remains close to $Q(t)\approx 1$, for $\ep, v$, which is expected in the case of shock formation in such $L_1$ convergence tests. We note also that the convergence order of elements of $T^{\mu \nu}$ showed $p >1 $ in Fig. \ref{fig9:ep_shock_convergence}.

While it is known that initial conditions like this result in shock formation in a similar fashion for ideal fluids \cite{ ChristodoulouShockDevelopment},  it is unclear what it means for a shock wave to form in BDNK theory; for a mathematical investigation of the existence and properties of shock profiles in conformal BDNK theory, see \cite{Freistuhler2021}. 
From the physics point of view, shock wave solutions formally feature diverging gradients and satisfy only the weak form of the hydrodynamic equations, but in a first-order theory like BDNK, this is somewhat problematic since gradients directly contribute to the energy-momentum tensor via the constitutive relations. Thus, it is not clear what shock formation means, and how much of the process can be taken to be physical and quantitative  in an effective theory constructed via a truncated derivative expansion such as BDNK; see Ref.~\cite{Pandya_2021} for a previous discussion of this point. However, if shocks can dynamically form in a consistent way in BDNK, this requires at least that the process should be robust with respect to the choice of causal and stable hydrodynamic frame.

\begin{figure}[htbp]
    \centering
    \includegraphics[width=1.0\textwidth]{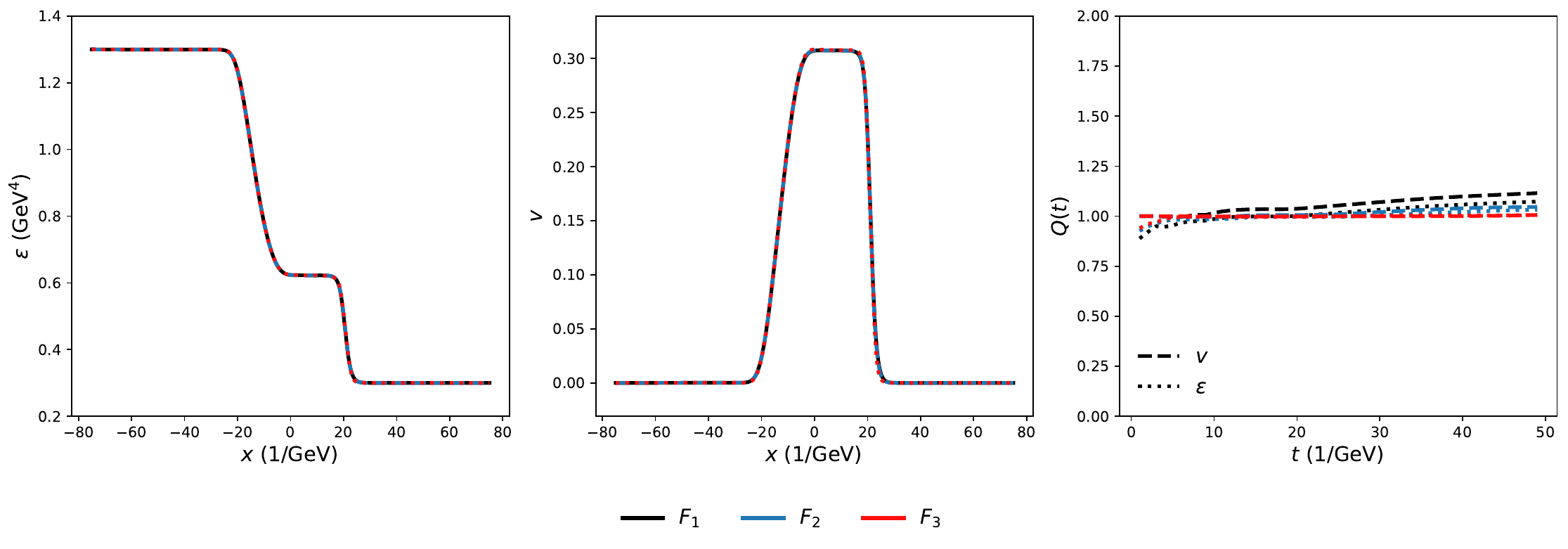}
    \caption{A snapshot of the hydrodynamic evolution at
    $t=30$ GeV$^{-1}$, alongside a $L_1$ convergence test  across all three hydrodynamic frames $F_1$, $F_2$ and $F_3$ in Table \ref{tab:hydro_frames},  for stair-step initial conditions, with  $4\pi \eta/s = 1$. The simulations pictured were performed on a grid of 48,001 cells.}
    \label{fig8:ep_shock_frames}
\end{figure}

The hydrodynamic frame dependence for the smooth step initial condition is studied in Fig.\ \ref{fig8:ep_shock_frames}, for both $\ep$, $v$, and their respective convergence functions. One can see that the profiles for the energy density and velocity are nearly identical for all frames considered. This suggests that the formation of such a sharp, shock-like structure in the dynamical evolution appears to be in the frame-robust regime of BDNK. 
Nevertheless, we remark that in the convergence test for $\ep$, we do see some difference between all frames. And we note that in $F_1$, the maximum characteristic speed is given by $c_+ = 1$, while frames $F_2$ and $F_3$ have subluminal maximum propagation speeds. Further work is needed to understand whether the sharp structures found here can indeed be considered as indicators of shocks. This, however, goes  beyond the scope of this first work, and we intend to return to this point in a future publication. Finally, we note that the time evolution of all components of $T^{\mu\nu}$, $C_\mu$, and $X_{\mu\nu}$ for initial conditions featuring a sharp (smooth) step in energy density can be found in Appendix \ref{appendix:gaussian}.

In the following, we use this smooth step in energy density initial data to provide a direct comparison between the viscous evolution of a BDNK fluid with that of an ideal fluid evolved using the relativistic Euler equations with a conformal equation of state. This is particularly interesting because such initial data lead to shock formation in the Euler case. 

\begin{figure}
    \centering
    \includegraphics[width=1\textwidth]{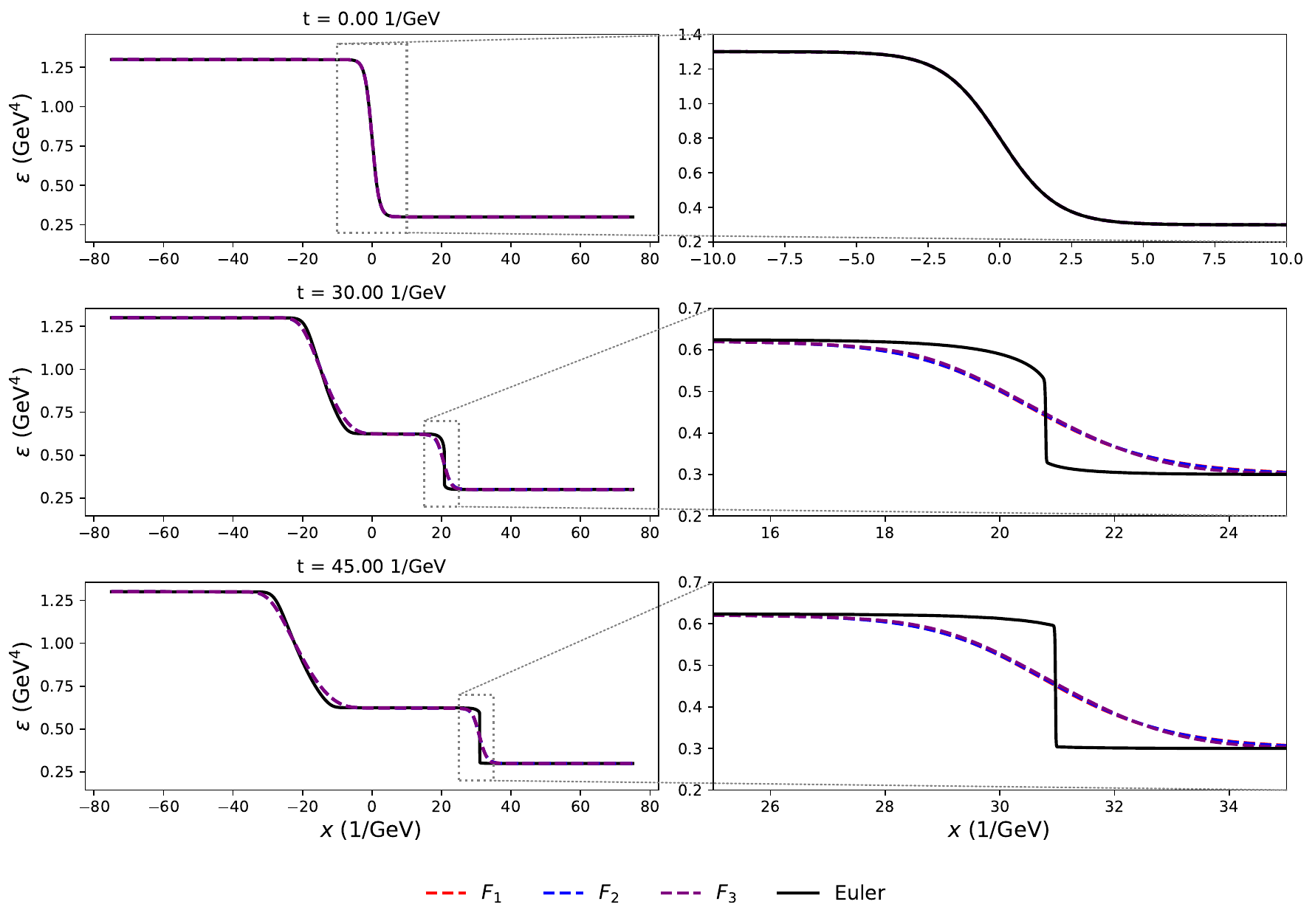}
    \caption{
    Comparing the hydrodynamic evolution between BDNK frames $F_1$, $F_2$ and $F_3$ in Table \ref{tab:hydro_frames} for viscosity at $4 \pi \eta/s = 1$, and an ideal fluid solution for the stair-step in energy density initial condition.}
    \label{fig10:BDNK_euler_shock}
\end{figure}

In Fig. \ref{fig10:BDNK_euler_shock}, we find evidence that there is initial data for which the dissipation in BDNK precludes shock formation in a frame robust manner. When comparing the evolution of a typical shock-inducing initial condition in Euler, we see the effects of viscous corrections on the structures of both the left (rarefaction wave) and right-moving (shock-wave) fronts. The right-hand side of this figure zooms in on what would be the shock-wave front in an ideal fluid, and we see a near-identical, though smooth structure, in the BDNK fluid across all the hydrodynamic frames used. This does not occur in an ideal fluid case, and we see this as evidence of the dissipative contributions in BDNK preventing shock formation in a  relativistic fluid.
This can occur because the BDNK equations are \emph{nonlinear}\footnote{To be more precise, the BDNK equations form a set of quasilinear hyperbolic PDEs \cite{AnileBook,Disconzi:2023rtt}. However, quasilinear PDEs can be nonlinear in the sense that the sum of solutions is not guaranteed to be a solution. This is the case, for example, for Euler's equations \cite{AnileBook}, and the same holds for BDNK.} hyperbolic equations.

Our results indicate that, depending on the initial data, the BDNK equations can prevent the formation of shocks:
a finite-speed, nonlinear regularization of compressive waves that
preserves finite-domain of dependence (hyperbolicity) while preventing gradient blow-up. This behavior---where hyperbolicity and regularization coexist---can be illustrated through a simpler hyperbolic model with linear damping, such as the damped version of the Burgers equation in 1+1 dimensions given by $\partial_t u +u\partial_x u =-\nu u$, where $\nu>0$, and $u=u(t,x)$ is some dynamical variable \cite{Rezzolla_Zanotti_book}; see Appendix \ref{appendix:burgers} for a detailed discussion about gradient blow up in this case. In fact, while the inviscid Burgers equation ($\nu=0$) forms shocks in finite time for any smooth initial data with $\partial_x u(x_0,t)<0$ at some point $x_0$, its dissipative counterpart can avoid this. If the initial data satisfy
\begin{align*}
\min_x \partial_x u(x,0) > -\nu,
\end{align*}
then the gradient $\partial_x u$ remains finite for all $t > 0$, and \emph{no shock forms}, as discussed in Appendix \ref{appendix:burgers}. In this regime, the solution is globally well-posed as the linear damping dominates over the nonlinear steepening, and for sufficiently small and smooth initial data (in the sense of a bounded gradient), dissipation successfully prevents shock formation, and the solution decays smoothly to zero in this case.

The same phenomenon---nonlinear smoothing of compressive gradients in the presence of finite propagation speeds---seems to be present in our BDNK simulations in the frame robust regime for the type of initial data considered. This, of course, does not rule out the presence of gradient blow-up in BDNK theory for other types of initial data, and further work is needed to investigate this question. Further analyses are clearly needed to systematically determine, both from a numerical and mathematical point of view, the fate of gradient blow-up in BDNK theory.

\section{Conclusions and Outlook} \label{section:Conclusion}

In this work we proposed a new first–order, flux–conservative formulation of relativistic BDNK hydrodynamics that is particularly well-suited  for shock–capturing methods. The formulation rewrites the dynamics in strict balance–law form with derivative–free sources in the evolved variables. This enables a straightforward semi–discrete central scheme with TVD time stepping and provides a practical, numerically robust pathway to probe sharp features without relying on the full characteristic decomposition. 

BDNK theory is a rigorous first–order framework in which small departures from equilibrium relax nonlinearly toward equilibrium in a manner consistent with relativity. Our work complements this by supplying a numerically ready, flux–conservative formulation together with operational diagnostics—convergence order, cross–frame agreement, and Knudsen proxies—that translate those mathematical guarantees into practical criteria for reliable simulations, and reveal how solutions may exit the regime of validity as gradients grow.

We have numerically studied the regime of validity of BDNK theory for two different initial problems in 1+1 dimensions: a smooth, stationary Gaussian profile and a sharply varying but smooth step profile that emulates a Sod–like setup. These problems were investigated across three admissible hydrodynamic frames, with the numerical evolution assessed by convergence testing. We considered an evolution to remain within the regime of validity when Knudsen numbers were small (i.e., $\ll 1$) and there was appreciable agreement across frames in both evolution and convergence, deeming these solutions robust to changes in hydrodynamic frame.

When a large initial viscosity was set for the Gaussian problem together with a relatively small minimum energy density, we encountered oscillations that are likely associated with the BDNK regulator fields $\mathcal A$ and $\mathcal Q^{\mu}$. This behavior is in agreement with previous results in the literature \cite{Pandya_2021,Bhambure:2024axa}. In our initial conditions (Sec.~\ref{section:IC}) these were set to vanish, but they are dynamically sourced by gradients. We verified whether runs remained in the regime of validity by monitoring normalized viscous contributions, using the Knudsen number as a practical proxy (Fig.~\ref{fig3:gaussian_Kn_visc}). At larger viscosities (set by $\eta/s$), these corrections were immediately driven away from zero and could become large on the next time step. We believe this behavior underlies the exotic evolution seen in the $\eta/s=20/4\pi$ case where regions appear where ${\rm Kn}\gg 1$. We showed that by simply increasing the minimum energy density at these larger viscosities, one can reduce gradients, stabilize the evolution, and return to the frame robust regime where the BDNK theory is reliable. While we use small Knudsen numbers as an operational indicator of validity, further studies are certainly needed as the precise boundary where the first–order theory breaks down is not yet known.

For the sharply varying step data, the time evolution showed visual agreement across frames, but the convergence test revealed to be a more sensitive probe: a reduced (first–order) $L_1$ convergence rate appeared for the step initial condition in Sec.~\ref{section:ICepshock}, together with the development of steep, right–moving structures. These are signatures typically associated with shock formation. We remark, however, that  
lower–order convergence is a necessary, though not sufficient, condition for shock formation in viscous, relativistic fluids, and was observed under some initial conditions for each problem (see Figs.~\ref{fig2:gaussian_convergence_delta0p1} and \ref{fig9:ep_shock_convergence}).

Prior work has shown that BDNK evolution can lead to shock–like structures starting from smooth initial data ~\cite{Pandya_2021,Freistuhler2021,Keeble:2025bkc}, and our results provide additional evidence in this direction, in the frame-robust regime, within a strictly flux-conservative formulation. Alongside this, we demonstrated evidence of frame-robust hyperbolic viscous dissipation overpowering shock formation in a relativistic fluid by comparing BDNK fluid evolution at low viscosity ($4\pi \eta/s = 1$) against an Euler fluid for the exact same initial state (see Fig. \ref{fig10:BDNK_euler_shock}). This is interesting because one expects that shocks form in Euler when compression exists, and what we find here is that, in BDNK, viscous effects can regularize incipient shocks for the class of small initial data we considered. This, however, does not preclude the formation of shocks and other singularities in BDNK for other types of initial data.

Our balance–law strategy is in the spirit of earlier efforts \cite{Pandya_2021}, but here we emphasize an explicit first–order reduction with derivative–free sources in the evolved variables, together with a consistency/equivalence narrative within the causal–stable set and a validation program focused on frame robustness and Knudsen diagnostics. Our focus has been flat–space $1{+}1$ dynamics with systematic hydrodynamic frame tests, which serves as a stepping stone toward realistic studies of the quark-gluon plasma formed in heavy-ion collisions \cite{Bea:2025eov}, and simulations in curved spacetimes (see \cite{Shum:2025jnl} for recent work in this regard).

In summary, the new flux–conservative formulation of BDNK hydrodynamics proposed here provides a stable and robust framework for studying viscous relativistic fluids. By testing both smooth and sharply varying initial data across multiple hydrodynamic frames, we established convergence properties and identified conditions under which solutions remain nearly frame–independent and physically reliable in the context of the first-order theory, while also observing signatures consistent with—though not conclusive of—shock formation.   These findings reinforce the potential of BDNK theory as a controlled setting to explore nonlinear, viscous fluid phenomena in relativistic  systems (e.g. quark-gluon plasma and neutron stars), and highlight the importance of monitoring  viscous corrections and frame dependence to ensure the reliability of the results.

Although our work was restricted to conformal fluids in $1{+}1$ Minkowski spacetime, the generality of the flux–conservative formulation proposed here paves the way for numerically investigating  nonconformal fluids, evolution scenarios beyond simple 1+1 dynamics, and curved spacetimes. Another particularly interesting direction is to develop a precise mathematical understanding of shock formation in BDNK hydrodynamics, for which our flux-conservative formulation may prove especially useful.

\section*{Acknowledgements}

We thank M.~Disconzi and J.~Anderson for insightful discussions about shocks in fluid dynamics. We also thank F.~Pretorius, A.~Pandya,  E.R.~Most, and P. Figueras for discussions concerning the numerical simulation of BDNK theory and H.~Witek for insightful discussions about convergence tests in numerical simulations. In particular, we thank A.~Pandya for providing a version of the code used in Ref.\ \cite{Pandya_2021}, which we used to check our solutions for the Gaussian initial data. J.N. thanks D.~Teaney and A.~Mazeliauskas for discussions concerning flux conservative formulations of hydrodynamics. J.N. and N.C. are partly
supported by the U.S. Department of Energy, Office of Science, Office for Nuclear Physics under Award No. DE-SC0023861.
E.P. is supported by Coordenação de Aperfeiçoamento de Pessoal de Nível Superior (CAPES), Grant No. 88887.838599/2023-00 and 88881.934515/2024-01, and working as part of the project Instituto Nacional de Ciência e Tecnologia—Física Nuclear e Aplicações (INCT—FNA), Grant No. 464898/2014-5. T.P. and F.S.B. gratefully acknowledge support from the Vanderbilt Initiative for Gravity, Waves, and Fluids, supported by Vanderbilt's College of Arts \& Sciences.
This work was partially done while the author F.S.B. was a Research Assistant Professor at Vanderbilt University.
M.H. was supported by the Brazilian National Council for Scientific and Technological Development (CNPq) under process No. 313638/2025-0.

\appendix 

\section{Results for the evolution of the fields in the flux conservative formulation}\label{appendix:gaussian}

\subsection{Gaussian in energy density}\label{section:appendix-ep-gaussian}

\begin{figure}[htbp]
    \centering
    \includegraphics[width=1.0\textwidth]{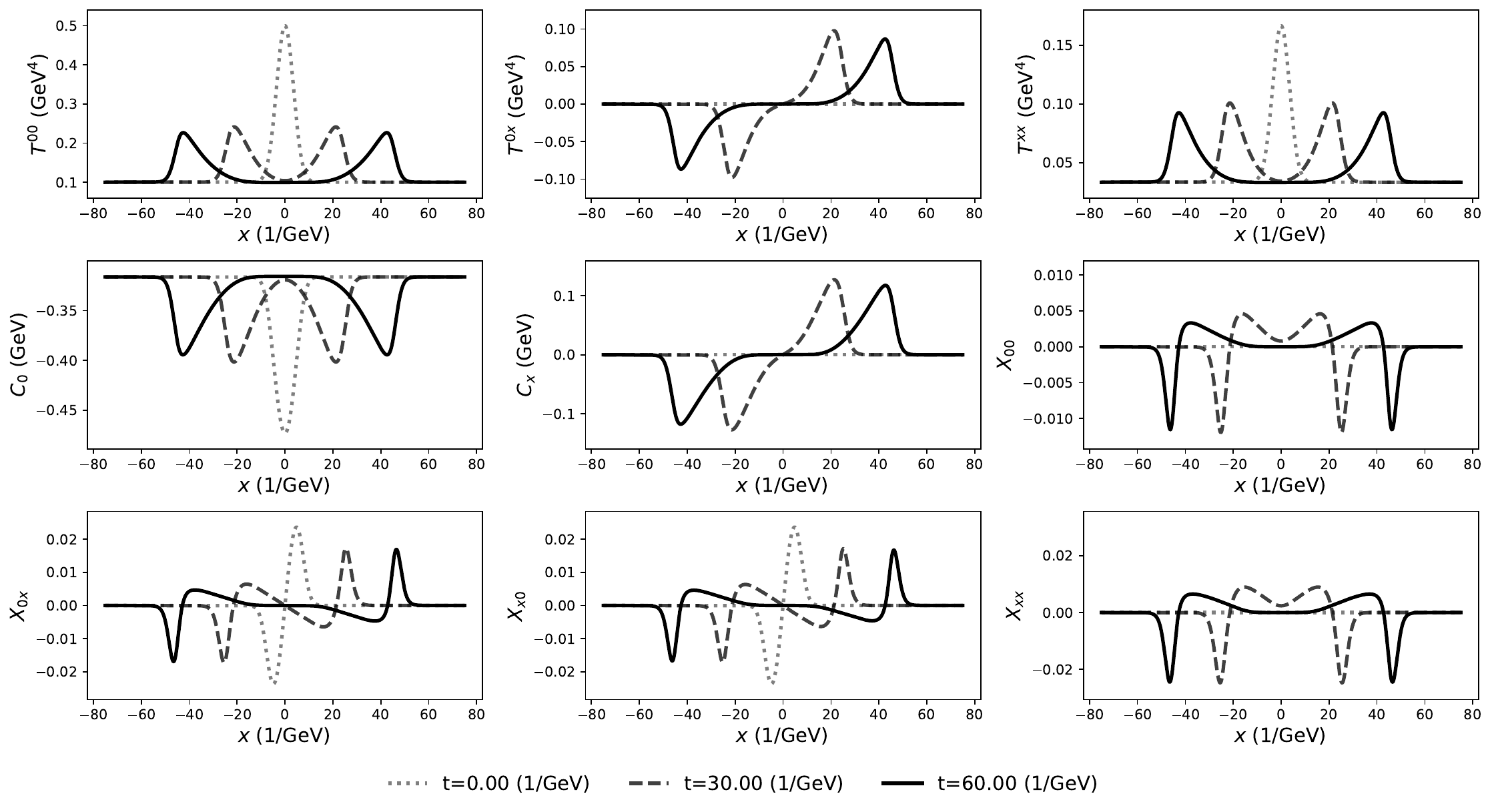}
    \caption{     Three snapshots of the evolution of all components of $T^{\mu\nu}$, $C_\mu$, and $X_{\mu\nu}$
     at
    $t=0,30,60$ GeV$^{-1}$,  for Gaussian initial conditions and $4\pi \eta/s = 1$. 
    Here we employ the $F_2$ hydrodynamic frame.
    The simulations were performed on a grid of 48,001 cells, with a minimum energy density of $\varepsilon_0 = 0.1$ GeV$^4$.}
    \label{fig11:gaussian_ep_visc}
\end{figure}

In Fig.~\ref{fig11:gaussian_ep_visc}, the evolution of all the evolved tensor and vector elements present in the state, flux, and source variables in \eqref{EOM1+1} is presented at three snapshots in time, $t= 0, 30, 60$ GeV$^{-1}$. Simulations were carried out for the initial data in Section \ref{section:ICgaussian}, with a viscosity set by $4\pi \eta/s = 1$ and a minimum energy density of $\varepsilon_0 = 0.1$ GeV$^4$, employing hydrodynamic frame $F_2$.

\subsection{Smooth step in energy density}\label{appendix:epshock}

\begin{figure}[htbp]
    \centering
    \includegraphics[width=1.0\textwidth]{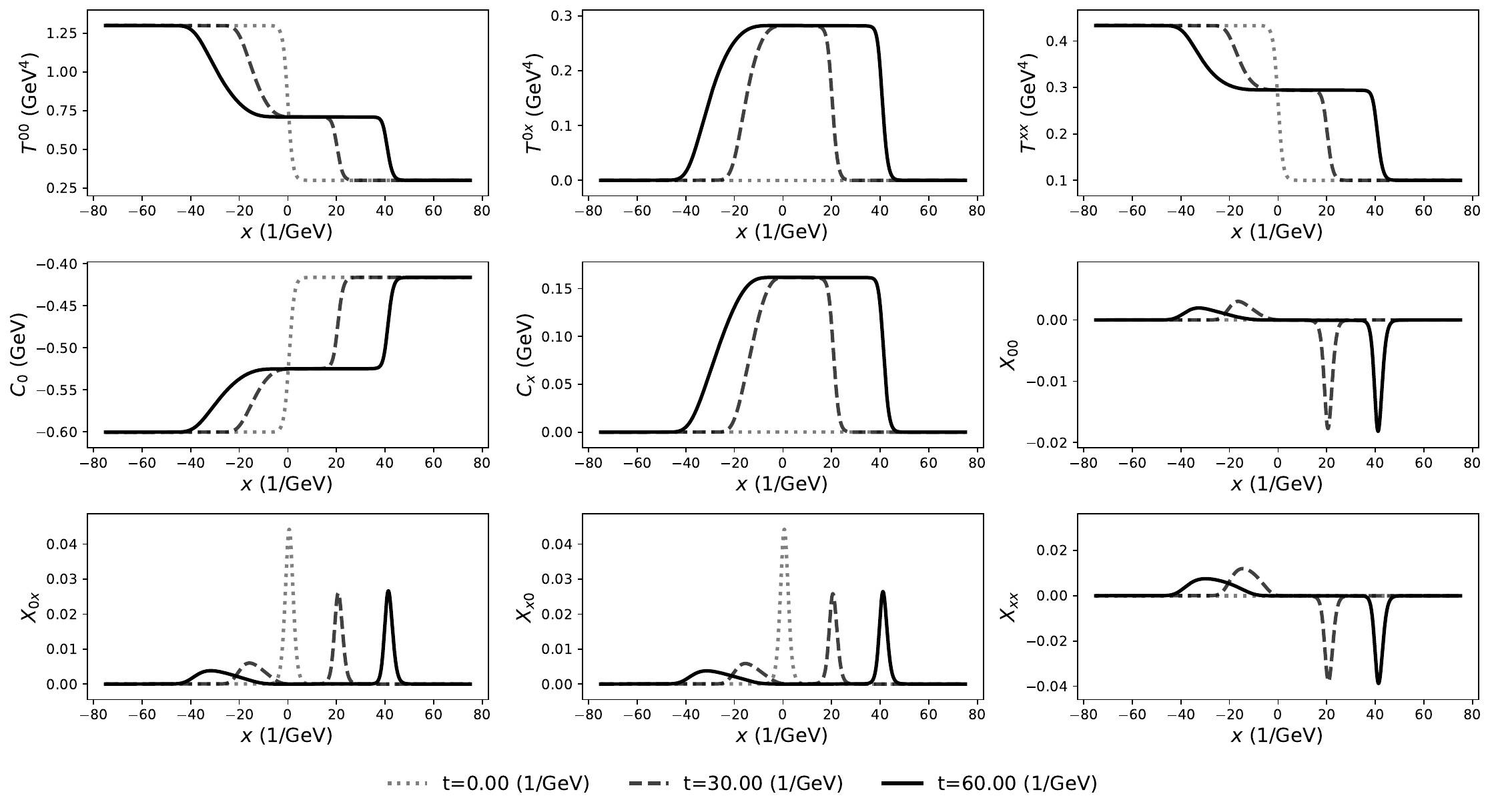}
    \caption{
     Three snapshots of the evolution of all components of $T^{\mu\nu}$, $C_\mu$, and $X_{\mu\nu}$
     at
    $t=0,30,60$ GeV$^{-1}$,  for stair-step initial conditions and $4\pi \eta/s = 1$. 
    Here we employ the $F_2$ hydrodynamic frame.
    The simulations were performed on a grid of 48,001 cells.
    }
    \label{fig12:shock_ep_visc}
\end{figure}

In Fig.~\ref{fig12:shock_ep_visc}, the evolution of all the evolved tensor and vector elements present in the state, flux, and source variables in \eqref{EOM1+1} is presented at three snapshots in time, $t= 0, 30, 60$ GeV$^{-1}$. This was done for the initial data in Section \ref{section:ICepshock}, with a viscosity set by $4\pi \eta/s = 1$, employing hydrodynamic frame $F_2$.

\section{Euler simulation details in 1+1 dynamics}
\label{appendix:Euler}

\begin{figure}[htbp]
    \centering
    \includegraphics[width=0.5\textwidth]{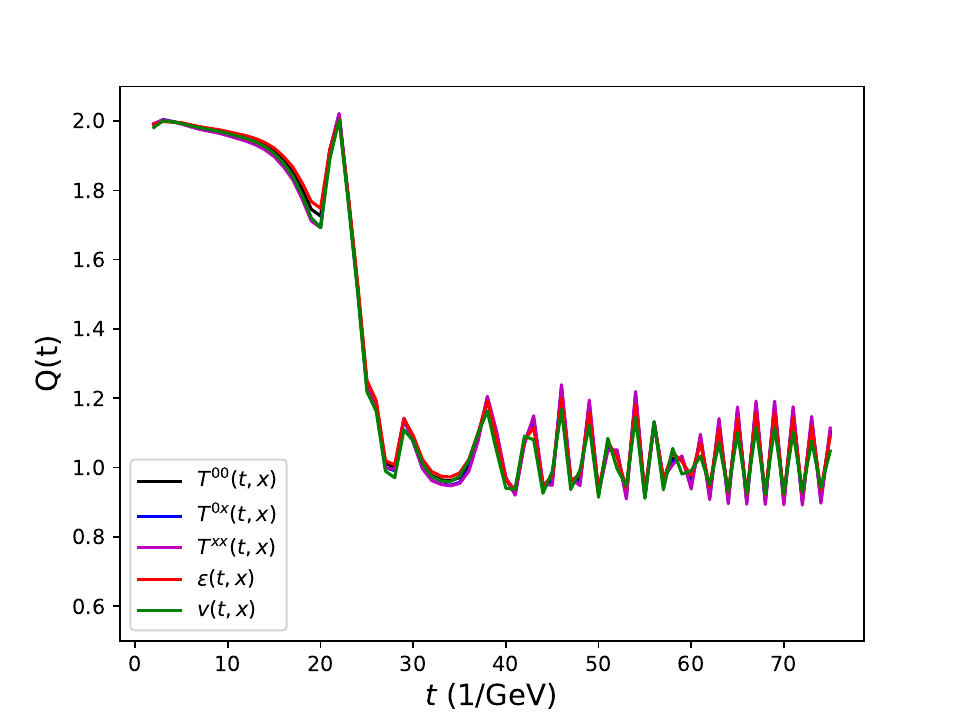}
    \caption{
     Convergence test for the Euler simulation without the use of any additional numerical dissipation to quell numerical oscillations present about the reduced convergence order $p \sim 1$, the testing was performed on grids of 6,001, 12,001, and 24,001 cells.
    }
    \label{fig13:euler_convergence}
\end{figure}

In flux conservative form, the relevant PDEs that define the 1+1 dimensional evolution of a relativistic ideal (Euler) fluid are \cite{Rezzolla_Zanotti_book} 
 \begin{align}
\partial_t E + \partial_x M &= 0,\\
\partial_t M + \partial_x S &= 0,
\end{align}
where $E\equiv T^{tt}_{0},\ \ M\equiv T^{tx}_{0},\ \ S\equiv T^{xx}_{0}$, and $T_0^{\mu\nu}$ is the ideal fluid energy-momentum tensor. Euler's equations are in flux conservative form, so we can use the same algorithms used to solve BDNK to perform the ideal fluid simulations needed in this work. 
As we advance in time from $t_l$ to $t_l+1$, we must have all the variables used to construct the flux function known. Using the relationship between the elements of $T_0^{\mu\nu}$, we find $S = Mv+P, E = M/v -P$ where ($u^x = \gamma v)$ and so
\eq{ S = Mv + M/v - E.}
We now have a flux-conservative equation that depends on the variables $E,M,v$. 
\eq{ \partial_t \m{ E \\ M \\  }
    + \partial_x \m{ M \\ \frac{M}{v} + Mv - E \\ } = \mathbf{0}.
}
We must be able to obtain $v$ from the evolved variables $E,M$ at each time step. Using the definition of $M$ and  the relation $\ep  =E- Mv$, we can obtain a general equation to solve that at first looks trivial to solve for $v$, but depending on the equation of state's complexity, one may run into polynomials with solutions that are roots for $v$,
\eq{ v(P+E) -M = 0.
}
Recall that, in general, the pressure is given by $P = P(\ep, ...)$. To construct the flux for all timesteps, we must have $\ep = E - Mv$, and so $P=P(v)$ at least. In the case of a linear equation of state $P=w\ep$ (for us we set $w=1/3$), we can obtain a quadratic to solve for $v$,
\eq{ v^2 w v_0 -v + v_0 = 0, \quad \text{where } v_0 = \frac{M}{E(w+1)}
}
with roots
\eq{ v_\pm = \frac{1}{2w v_0} \left ( 1 \pm \sqrt{1 - 4 wv_0^2} \right ).
}
Of the two possible roots, the one with the minus sign ($-$) is the only physical one, as we can recover the non-relativistic limit, $v_0 = v_{nr}$, whereas by using the other root, the velocity diverges. It is necessary to solve this equation at every time step in order to calculate the flux. We can rationalize the solution for $v_-$ by using the formula 
\eq{
1 - \sqrt{1-\delta} = \frac{\delta }{1 + \sqrt{1-\delta}},
}
thereby making the smallest root
\eq{
v_{-} = \frac{2v_0}{1 + \sqrt{1 - 4 w v_0^2} }.
}
Carrying out this procedure allows us to avoid numerical errors related to floating-point precision in the practical implementation of this expression in the simulation of the ideal fluid. The convergence test for the Euler simulation used in this work is found in Fig. \ref{fig13:euler_convergence}. The variations around the expected convergence rate $p \sim 1$, are typical for problems involving shock formation. In fact, solving shocks in relativistic Euler equations is known to produce spurious oscillations around the shock front. Methods like Kreiss-Oliger dissipation \cite{kreiss1973methods,Rezzolla_Zanotti_book} are commonly employed to suppress these spurious oscillations, but we do not employ any such additional dissipation throughout this work.

\section{The damped hyperbolic Burgers equation}
\label{appendix:burgers}

A key feature of first-order formulations of dissipative hydrodynamics, such as the BDNK theory, is that it can be reduced to a set of first-order dynamical equations, as shown in the main text, with the particularity that the principal part is first order in derivatives, while the dissipation enters through a source term that depends only on the dynamical variables, not their derivatives. In 1+1 dimensions, such a system can be written as $\partial_t \mathbf{q} + A \partial_x \mathbf{q} = \mathbf{S}(\mathbf{q})$, where $A_{ab} = \partial F_a / \partial q_b$. To isolate and study the mechanism by which dissipation can compete with nonlinear steepening in such hyperbolic systems, it is instructive to consider a simplified model. A suitable candidate is the damped Burgers equation
\begin{align}
\label{Burger_1}
\partial_t u + u \, \partial_x u = -\nu u, \quad \nu > 0.
\end{align}
This equation provides a clear testing ground for whether dissipation can prevent shock formation from smooth initial data, a process that is inevitable in the inviscid case ($\nu = 0$). The nonlinear term $u \partial_x u$ is the mechanism responsible for gradient blow-up in the inviscid Burgers equation \cite{Rezzolla_Zanotti_book}. Although the presence of analogous nonlinear modes in the BDNK formulation is a complex issue, the damped Burgers equation \eqref{Burger_1} offers valuable insight into how first-order \emph{hyperbolic} systems with dissipation can prevent the evolution to discontinuous solutions present in the inviscid limit.

To analyze the effect of dissipation, we examine solutions along characteristic curves $x(t)$, defined by $\dfrac{dx}{dt} = u(x(t),t)$. Along these curves, Eq. \eqref{Burger_1} becomes
\begin{align*}
\frac{du}{dt} = \partial_t u + u \, \partial_x u = -\nu u \quad \Rightarrow \quad u(t) = u_0 e^{-\nu t},
\end{align*}
where $u_0 = u(x(0),0)$. This shows that the amplitude of $u$ decays exponentially along characteristics. While this damping is significant, it does not, by itself, guarantee that shock formation is prevented.

To better understand the nature of the damping term $-\nu u$, define the energy 
$E(t) = (1/2) \int u^2 \,dx$
(with periodic or decay boundary conditions). Multiplying Eq.\ \eqref{Burger_1} by $u$ and integrating over $x$ yields
\begin{align*}
\frac{dE}{dt} = -2\nu E \quad \Rightarrow \quad E(t) = E(0)e^{-2\nu t}.
\end{align*}
Thus, we see that the term $-\nu u$ causes exponential energy decay, confirming its dissipative nature. However, this dissipation acts by reducing amplitudes while keeping the equation hyperbolic. 

Nevertheless, this term does not preclude shock formation --- it provides no gradient smoothing. As we show below, shocks can still develop if initial gradients are sufficiently steep, though they may be avoided for sufficiently small and smooth initial data. This contrasts with the effects from a diffusion term $\nu \partial_{x}^2 u$, present in the \emph{parabolic} Burgers equation $\partial_t u + u \partial_x u = \nu \partial_x^2 u$, which dissipates energy via gradients and provides parabolic regularization (the equation ceases to be hyperbolic) and can prevent shock formation for $\nu > 0$ \cite{DafermosConservationLawsBook}.

To study the potential for gradient blow-up, we define $v = u_x=\partial_x u$ and differentiate \eqref{Burger_1} with respect to $x$ to obtain
\begin{align*}
v_t + u v_x + v^2 = -\nu v.
\end{align*}
Following this relation along a characteristic curve yields
\begin{align*}
\frac{dv}{dt} = -v^2 - \nu v,
\end{align*}
where, along the characteristic curve, $v=v(x(t),t)$ and $d/dt=\partial_t+u(x(t),t)\partial_x$.
This Riccati equation has the exact solution
\begin{align*}
v(t) = \frac{\nu v_0 e^{-\nu t}}{v_0 (1 - e^{-\nu t}) + \nu}, \quad v_0 = v(0).
\end{align*}
A finite-time gradient blow-up (shock formation) occurs if $v(t) \to -\infty$, which happens when the denominator vanishes. This condition is
\begin{align*}
v_0 (1 - e^{-\nu t_c}) + \nu = 0.
\end{align*}
Solving for the critical time $t_c$ gives $1 - e^{-\nu t_c} = -\dfrac{\nu}{v_0}$. Since the left-hand side lies in the interval $[0, 1)$, a necessary condition for blow-up is
\begin{align*}
0 \le -\frac{\nu}{v_0} < 1.
\end{align*}
Given that $v_0 < 0$ is required for wave steepening, this inequality implies the crucial condition
\begin{align*}
v_0 < -\nu.
\end{align*}
Therefore, shocks form only if the initial negative slope $v_0$ is more negative than $-\nu$.

Conversely, if the initial data satisfy
\begin{align*}
\min_x u_x(x,0) > -\nu,
\end{align*}
then the denominator never vanishes, the gradient $v= \partial_x u$ remains finite for all $t > 0$, and \emph{no shock forms}. In this regime, the solution is globally well-posed. The linear damping dominates over the nonlinear steepening, and for sufficiently small and smooth initial data (in the sense of a bounded gradient), dissipation successfully prevents shock formation, and the solution decays smoothly to zero.

Regardless of the specific structure of nonlinear modes in the first-order reduction of BDNK theory, this analysis of the dissipative Burgers equation illustrates a fundamental mechanism: in first-order hyperbolic systems (which, in our context, arise from a first-order reduction of second-order PDEs), dissipation can preclude shock wave formation from smooth, small initial data, which is sharp contrast with  dissipation-free hyperbolic systems such as the inviscid Burgers or Euler equations.

\section{Convergence testing framework} 
\label{appendix:convergence}

Convergence testing provides a systematic procedure for verifying that numerical solutions approach the continuum limit as the grid spacing $h$ is refined. The central idea is that a consistent and stable discretization of order $p$ should yield an error that decreases as a power of the grid spacing, $| u_h - u | \sim h^p$, where $u_h$ denotes the numerical solution obtained on a mesh with spacing $h$ and $u$ represents the continuum solution. In practice, since exact solutions are rarely available for nonlinear systems such as relativistic hydrodynamics, convergence is verified by comparing simulations performed at different resolutions, and by measuring how the differences between them scale with grid refinement. A numerical method that exhibits the expected scaling is regarded as convergent within the tested regime.

This scaling behavior can be formally justified through Richardson expansion. In this framework, the numerical approximation is expressed as
\begin{equation}
A(h) = A + c h^p + \mathcal{O}(h^{p+1}),
\end{equation}
where $A$ is the exact solution, $p$ is the order of accuracy, and $c$ is a constant that depends on the discretization and problem setup. The leading error term is proportional to $h^p$, and successive refinements of the mesh reduce this contribution by a predictable factor. Considering three grid spacings related by $h_c = r h_m$ and $h_m = r h_f$, where the subscripts $c,m,f$ refer to coarse, medium, and fine grid spacing, the effective convergence rate may be extracted directly from ratios of solution differences,
\begin{equation}
p_{\text{eff}} = \log_r \left( \frac{| q_{h_c} - q_{h_m} |}{| q_{h_m} - q_{h_f} |} \right).
\end{equation}
This expression follows immediately from the Richardson form of the error and provides a practical diagnostic for verifying whether a numerical implementation achieves its theoretical order of accuracy under grid refinement \cite{stoer2002introduction}.

In this study, we evaluate the convergence of our numerical implementation of the flux-conservative formulation of relativistic viscous hydrodynamics, as described by the BDNK theory. To quantify differences between grid levels, we employ an $L_1$ norm applied to the hydrodynamic fields and components of the stress tensor. The choice of the $L_1$ norm is motivated by its robustness: it provides a global measure of the integrated error across the computational domain, it is less sensitive than the $L_\infty$ norm to isolated pointwise deviations, and it offers a clearer interpretation in the presence of shocks or sharp gradients than the $L_2$ norm. 
The convergence function is then defined as
\begin{equation} \label{convergence_test}
Q(t) = \log_2 \left (\frac{ | q^i(x; h_c) - q^i(x; h_m) |_1}{| q^i(x; h_m) - q^i(x; h_f) |_1} \right),
\end{equation}
where the discrete $L_1$ norm is computed over the spatial grid at a fixed coordinate time and we take $r=2$. This ratio directly measures the rate at which the numerical error decreases under successive resolution doubling.

The simulations use the Kurganov–Tadmor central scheme for spatial discretization (formally second-order accurate in smooth regions) together with a second-order Runge–Kutta method for time integration.
Accordingly, in the convergent regime we expect $Q=2$.
Monitoring this quantity over the course of the evolution provides a quantitative check that the implementation is performing at the expected order of accuracy. Deviations from the theoretical limit may indicate loss of resolution, the presence of shocks or discontinuities, or other numerical artifacts that reduce the effective order of the scheme.


\section{Comparison to previous work}
\label{appendix:pandya_agreement}

\begin{figure}[htbp]
    \centering
    \includegraphics[width=0.7\textwidth]{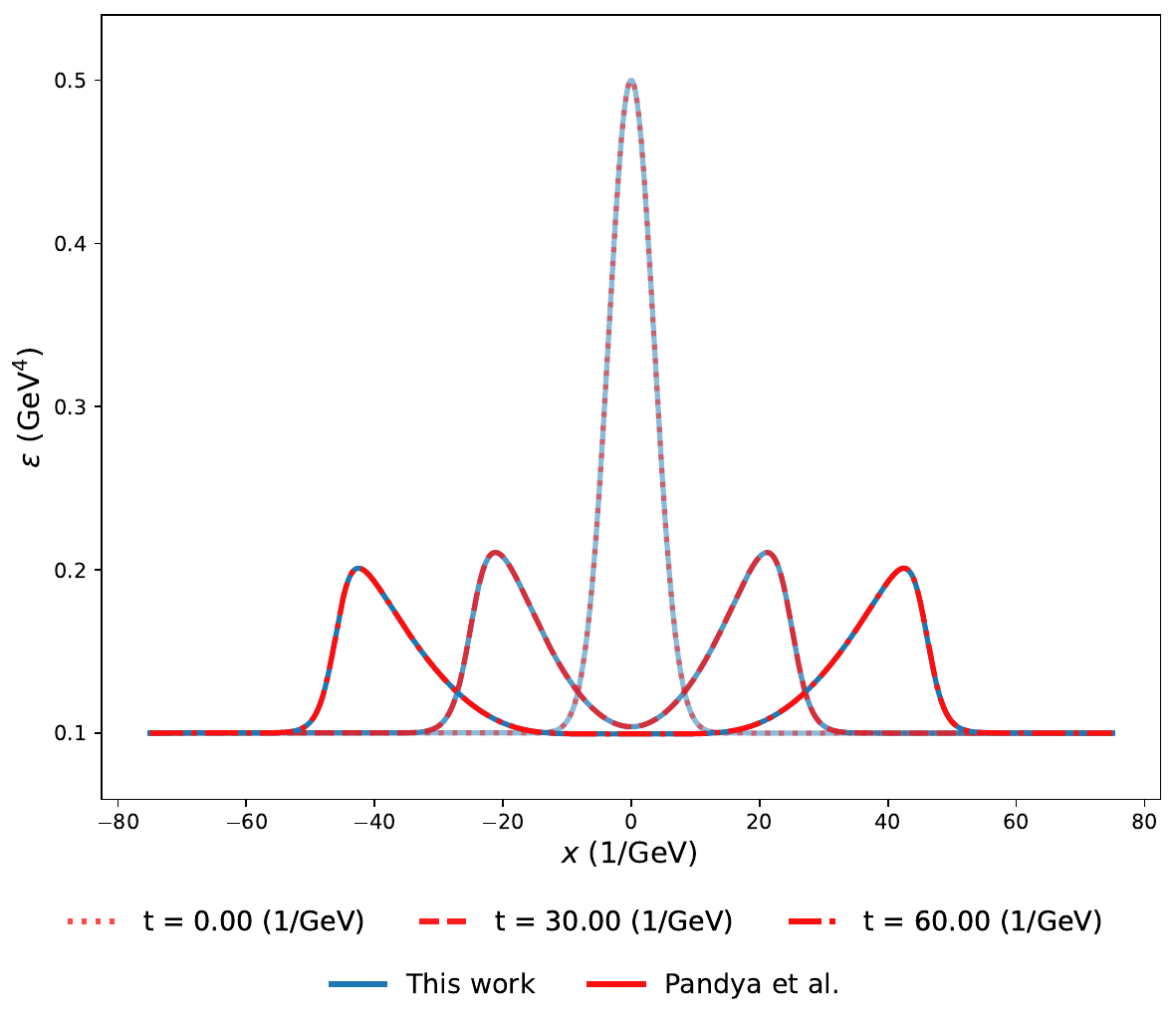}
    \caption{
    BDNK simulation agreement between this work and the work of Ref.\ \cite{Pandya_2021}.
    }
    \label{fig14:pandya_agreement}
\end{figure}

Here we draw comparisons between our work and the work of Ref.\ \cite{Pandya_2021} for the case $\eta/s = 1/4\pi$ of the Gaussian problem explained in Sec. \ref{section:ICgaussian}. We see complete agreement between our two formulations in Fig. \ref{fig14:pandya_agreement} where we used the same hydrodynamic frame $F_1$ outlined in Table \ref{tab:hydro_frames}.

\def\cprime{$'$}

\end{document}